\documentclass[a4paper,12pt]{article}
\usepackage{amssymb,amsthm,chicago,booktabs,rotating,threeparttable}
\usepackage{algorithm,algorithmicx,algpseudocode}
\usepackage{graphicx,subfigure}
\usepackage{amsmath}
\usepackage{setspace}
\usepackage{lscape}
\usepackage[textwidth=6.1in,centering]{geometry}
\usepackage{lineno}

\onehalfspace

\newcommand{\bq}[1]{\begin{equation}\label{#1}}
\newcommand{\eq}{\end{equation}}
\newcommand{\bqn}{\begin{eqnarray}}
\newcommand{\eqn}{\end{eqnarray}}
\newcommand{\bqns}{\begin{eqnarray*}}
\newcommand{\eqns}{\end{eqnarray*}}

\newcommand{\eps}{\varepsilon}

\newcommand{\dd}{\,\textrm{d}}

\newcommand{\Nn}{\textrm{\rm N}}

\def\eps{\varepsilon}
\def\({\left(}
\def\){\right)}

\begin{document}

\title{Particle Efficient Importance Sampling}
\author{Marcel Scharth  \ \ Robert Kohn\\
\small{
Australian School of Business, University of New South Wales }}
\date{\small September 25, 2013}

\maketitle

\vspace{-8mm}
\begin{abstract}
\noindent The efficient importance sampling (EIS) method is a general principle for the numerical evaluation of high-dimensional integrals that uses the sequential structure of target integrands to build variance minimising importance samplers.  Despite a number of successful applications in high dimensions, it is well known that importance sampling strategies are subject to an exponential growth in variance as the dimension of the integration increases. We solve this problem by recognising that the EIS framework has an offline sequential Monte Carlo interpretation. The particle EIS method is based on non-standard resampling weights that take into account the look-ahead construction of the importance sampler. We apply the method for a range of univariate and bivariate stochastic volatility specifications. We also develop a new application of the EIS approach to state space models with Student's $t$ state innovations. Our results show that the particle EIS method strongly outperforms both the standard EIS method and particle filters for likelihood evaluation in high dimensions. Moreover, the ratio between the variances of the particle EIS and particle filter methods remains stable as the time series dimension increases. We illustrate the efficiency of the method for Bayesian inference using the particle marginal Metropolis-Hastings and importance sampling squared algorithms.

\footnotesize{

}
\end{abstract}

\newpage


\section{Introduction}

This paper introduces the particle efficient importance sampling (P-EIS) method as a tool for likelihood evaluation and state inference in nonlinear non-Gaussian state space model applications. The approach is based on the EIS algorithm of \citeN{ZR2007}, which is an importance sampling method for the estimation of high-dimensional integrals that have a sequential structure. The EIS method constructs global approximations to target integrands by iterating a sequence of least-squares regressions, which are linear and therefore computationally efficient for a wide range of models. The essential idea of the P-EIS method is that the high-dimensional EIS approach has a sequential Monte Carlo (SMC) interpretation compatible with the introduction of resampling steps. We show that it is crucial to use non-standard resampling weights that take into account the look-ahead construction of the importance sampler.

The use of importance sampling to evaluate the likelihood of nonlinear non-Gaussian state space models for long time series dates back to the method of \citeN{SP97} and \citeN{DK97}, which relies on a Laplace approximation to the likelihood. The use of the global approximation technique in the EIS method has expanded the scope of high-dimensional importance sampling, and a range of applications are now available in the literature. Some examples include stochastic volatility models in \citeN{LR2003}, the stochastic conditional duration model in \citeN{BG09}, probit models with correlated errors in \citeN{lr2010}, DSGE models in \citeN{dlmrd2012}, stochastic copula models in \citeN{hm2012}, state space models with mixture measurement densities in \citeN{kl2013}, discrete dependent variable models with spatial correlation in \citeN{lrv2013}, and the corporate default model in \citeN{bhkl2013}.

Despite these successful applications, the use of importance sampling has so far been limited by the exponential increase in the variance of the likelihood estimate as the dimension of the integration problem increases. See for example \citeN{chopin2004}, Section 3.3. The SMC approach of the particle EIS method solves this problem by introducing resampling when generating draws from a high-dimensional importance density.  That also includes the method of \citeN{SP97} and \citeN{DK97}, which \citeN{NAIS} show to be compatible with the sequential implementation of the EIS sampler. Like the standard EIS method, the particle EIS algorithm aims to explicitly minimise the variance of the likelihood estimate using all the available sample information. The same is typically not the case with particle filters, which are limited by design to focus on conditional optimality: minimising the variance of the importance weights in the current period, given the particles propagated from the previous period. We show how the particle EIS method also directly addresses the numerical inefficiency that particle filters are subject to even if the conditionally optimal (fully adapted) proposal is feasible.

The particle EIS method belongs to the class of auxiliary particle filters (APF) introduced by \citeN{ps1999}. It therefore provides an unbiased estimate of the likelihood following the general result for auxiliary particle filters in \citeN{delmoral2004}. This property is fundamental for  applications to Bayesian inference using the particle marginal Metropolis-Hastings (PMMH) method of \citeN{pmcmc} and the importance sampling squared (IS$^2$) method of \citeN{tspk2013}. See also the discussion in \citeN{fs2011}.

We present a detailed study of the numerical efficiency of the particle EIS method compared to the EIS algorithm and standard particle filters. We base our analysis on a simulation study for a range of univariate stochastic volatility (SV) models and a bivariate SV specification, for which we also present an empirical application. Our general univariate specification allows for a fat-tailed measurement density, a two-factor log-volatility process, leverage effects (which imply a nonlinear state transition) and additive Student's $t$ state innovations, highlighting the flexibility of the EIS framework. The application of EIS for models with additive Student's $t$ state disturbances is new to the literature. We develop the EIS algorithm for this model using a data augmentation idea initially proposed by \citeN{kl2013}.

The simulation study leads to three main conclusions. First, the particle EIS method brings large reductions in variance over the standard EIS method. For a time series of $10,000$ observations, the decrease in variance ranges from $80\%$ for the univariate SV model with Student's $t$ state disturbances to 95\% for the bivariate specification. These gains come with a negligible increase in computational time. Second, the EIS and P-EIS methods strongly outperform standard particle filters for these models. Our result show that the P-EIS method outperforms the best particle filter in our analysis by factors of approximately 100 to 6,000, depending on the specification under consideration. Third, the particle EIS method approximately maintains a constant performance relative to the particle filters for all time series dimensions.

The empirical application for the bivariate SV model uses $5,797$ daily observations of the IBM and General Electric stock returns. We focus on posterior inference using the PMMH and IS$^2$ methods. Using the theory on the optimal implementation of these two methods developed by \citeN{psgk2012} and \citeN{tspk2013} respectively, we find that for this example the P-EIS method needs only 10 particles to achieve the same numerical performance for posterior inference as a bootstrap filter with 15,000 particles. This result shows that particle EIS can make Bayesian estimation of complex state space models feasible in situations in which simple particle methods require unreasonable computing times for accuracy. As in the simulation study, our empirical analysis shows substantial gains from using the P-EIS method in comparison with the standard EIS algorithm.

The use of the EIS principle to address the limitations of particle filters has also been considered by \citeN{dlmrd2012}. In that paper, the authors introduce the EIS filter for likelihood evaluation in state space models applications. Their method consists of using the EIS method to construct continuous approximations of filtering densities that result in unconditionally optimal approximations of target integrands. Our method represents a distinct approach. We maintain the focus in approximating the smoothing density of the states as in \citeN{ZR2007}, in contrast with the approximation of filtering densities in the EIS filter. The EIS filter represents a different approach to particle methods and does not entail resampling. Finally, the EIS filter estimate of the likelihood is biased but continuous, whereas the particle EIS estimate is unbiased but discontinuous. In this sense, we can view the particle EIS method and the EIS filter as complementary approaches as unbiasedness and continuity are relevant properties for Bayesian and classical estimation respectively.

Our method additionally relates to previous contributions on look-ahead and block sampling strategies for sequential Monte Carlo, see for example \citeN{lcl2013} and \citeN{dbs2006} on these two topics respectively.  We can view the particle EIS method as a generalisation of some of these ideas that allows for the construction of an importance density that incorporates all available information into a high-dimensional sampler, which we break into smaller blocks as resampling becomes appropriate. It is straightforward to modify the method to specialised settings that use partial information and smaller block sizes.

We organise the paper as follows. Section \ref{sec:IS} presents the notation and estimation objective and reviews the EIS method. Section \ref{sec:PEIS} introduces and motivates the particle EIS method. Section \ref{sec:simulation} studies the relative performance of the new method for likelihood evaluation for univariate and bivariate stochastic volatility models in a simulated setting. Section \ref{sec:empirical} presents an empirical application to posterior inference via IS$^2$ and PMMH.

\section{Importance sampling}\label{sec:IS}

\subsection{State Space Model}

Consider a discrete-time Markov process $\{X_t\}_{t\geq1}$ such that
\[X_1\sim p(x_1),\qquad X_t|(X_{t-1}=x_{t-1})\sim p(x_t|x_{t-1}).\]
We assume that $n$ observations are generated by the measurement density
\[Y_t|(X_{t}=x_{t})\sim p(y_t|x_{t}).\]
The state and measurement densities implicitly depend on a parameter vector $\theta\in \Theta \subseteq \mathbb{R}^d$, which we omit from the notation whenever possible for conciseness. Define $x_{1:t}=(x_1' \, , \, \ldots \, , \, x_t')'$ and $y_{1:t}=(y_1' \, , \, \ldots \, , \, y_t')'$. The likelihood for the state space model is given by the integral
\begin{align}\label{lik}
    L(y_{1:n})&=\int p(y_{1:n},x_{1:n})\dd x_{1:n}=\int p(y_{1:n}|x_{1:n})p(x_{1:n})\dd x_{1:n}\\
    &=\int p(y_1|x_1)p(x_1) \prod_{t=2}^{n} p(y_t|x _t)p(x_t|x_{t-1})\dd x_1\ldots \dd x_n,\nonumber
\end{align}
which is typically analytically intractable. Our objective in this paper to obtain an accurate and unbiased Monte Carlo estimate $\widehat{L}(y)$ of this integral for a wide class of models.

\subsection{Efficient High Dimensional Importance Sampling}\label{sec:EIS}

To evaluate the likelihood function by importance sampling, we consider a high-dimensional importance distribution $q(x_{1:n}|y_{1:n})$ and rewrite the likelihood function as
\begin{eqnarray}\label{basicid}
L(y_{1:n}) & = & \int \frac{p(y_{1:n}|x_{1:n})p(x_{1:n})}{q(x_{1:n}|y_{1:n})}q(x_{1:n}|y_{1:n})\dd x_{1:n} \\
     & = &\int \omega(x_{1:n},y_{1:n}) q(x_{1:n}|y_{1:n})\dd x_{1:n}, \nonumber
\end{eqnarray}
where the importance weight function is given by
\begin{equation}\label{isweight}
\omega(x_{1:n},y_{1:n})=\frac{p(y_{1:n}|x_{1:n})p(x_{1:n})}{q(x_{1:n}|y_{1:n})}.
\end{equation}

We estimate the likelihood function \eqref{lik} by generating $N$ independent trajectories $x_{1:n}^{(1)} \, , \, \ldots \, , \, x_{1:n}^{(N)}$
from the importance density $q(x_{1:n}|y_{1:n})$ and computing
\begin{equation*}\label{likest}
\widehat{L}(y_{1:n})=\bar \omega, \qquad
\bar \omega = \frac{1}{N}\sum_{i=1}^{N} \omega_i, \qquad
\omega_i = \omega(x_{1:n}^{(i)},y_{1:n}),
\end{equation*}
where $\omega _i$ is the realised importance weight function in \eqref{isweight} for $x_{1:n} = x_{1:n}^{(i)}$. \citeN{Geweke89} showed that a central limit theorem applies to the importance sampling estimate provided that
\begin{eqnarray*}
     \int \omega(x_{1:n},y_{1:n})^2 q(x_{1:n}|y_{1:n})\dd x<\infty,
\end{eqnarray*}
in which case the estimate is asymptotically normal and converges at the regular parametric rate to the true likelihood. A sufficient condition for the integral above to be finite is that the importance weight function is bounded from above. \citeN{KSC09} used extreme value theory to develop diagnostic tests to validate the existence of the variance of the importance weights.

The high-dimensional efficient importance sampling method of \citeN{ZR2007} considers an importance sampler with the following form
\begin{equation*}
q(x_{1:n}|y_{1:n})=q(x_1|y_{1:n})\prod_{t=2}^{n}q(x_{t}|x_{t-1},y_{1:n}).
\end{equation*}

It follows that the we can factorise the importance weight as
\begin{equation}\label{isweight2}
\omega(x_{1:n},y_{1:n})=\frac{p(y_{1}|x_{1})p(x_{1})}{q(x_{1}|y_{1:n})}\prod_{t=2}^{n}\frac{p(y_{t}|x_{t})p(x_{t}|x_{t-1})}{q(x_{t}|x_{t-1},y_{1:n})}.
\end{equation}

\citeN{ZR2007} write the conditional densities $q(x_{t}|x_{t-1},y_{1:n})$ in terms of a kernel in $x_t$ and an integration constant
\begin{equation}\label{kernel}
q(x_{t}|x_{t-1},y_{1:n})=\frac{k(x_t,x_{t-1};a_t)}{\chi(x_{t-1};a_t)},
\end{equation}
where
\begin{equation}
\chi(x_{t-1};a_t)=\int k(x_t,x_{t-1};a_t) \dd x_t
\end{equation}
and $a_t$ is a vector of importance parameters which depends on $y_{1:n}$. At the initial period, we have the density
\[q(x_{1}|y_{1:n})=\frac{k(x_1;a_1)}{\chi(a_1)},\qquad \chi(a_1)=\int k(x_1;a_1) \dd x_1.\]

Using \eqref{isweight2} and \eqref{kernel}, we express the importance sampling identity \eqref{basicid} as
\begin{eqnarray} \label{basicid2}
     &   & \int \frac{p(y_{1}|x_{1})p(x_{1})}{q(x_{1}|y_{1:n})} \prod_{t=2}^{n}\frac{p(y_{t}|x_{t})p(x_{t}|x_{t-1})}{q(x_{t}|x_{t-1},y_{1:n})}q(x_{1:n}|y_{1:n})\dd x_{1:n}\\
         & = &\chi(a_{1}) \int \frac{p(y_{1}|x_{1})p(x_{1})\chi(x_{1};a_{2})}{k(x_1;a_1)}\prod_{t=2}^{n}\frac{p(y_{t}|x_{t})p(x_{t}|x_{t-1})\chi(x_{t};a_{t+1})}{k(x_t,x_{t-1};a_t)} q(x_{1:n}|y_{1:n})\dd x_{1:n}, \nonumber
\end{eqnarray}
with the convention that $\chi(x_{n};a_{n+1})\equiv1$.

The EIS method seeks to find importance parameters $a_t$ which minimise the variance of the ratio
\begin{eqnarray}\label{ratio}
    \frac{p(y_{t}|x_{t})p(x_{t}|x_{t-1})\chi(x_{t};a_{t+1})}{k_t(x_t,x_{t-1};a_t)},
\end{eqnarray}
where the backward shifting of the period $t+1$ integration constant $\chi(x_{t};a_{t+1})$ is essential for obtaining a numerically efficient estimate of the joint integral \eqref{lik}. This is intuitive given the dependence of the integration constant on the lagged state. \citeN{KLS3} note that when both the measurement and transition densities are linear Gaussian, letting $k_t(x_t,x_{t-1};a_t)\propto p(y_{t}|x_{t})p(x_{t}|x_{t-1})\chi(x_{t};a_{t+1})$ leads to an analytical backward-forward smoother and an efficient simulation smoother for this class of models, with the likelihood being computed exactly as a side product.

\citeN{ZR2007} propose Algorithm \ref{alg1} for selecting the importance parameters $a_{1:n}$. We highlight some critical aspects of it. The use of common random numbers (CRN) ensure the smoothness of the criterion function across successive iterations, facilitating the convergence of the algorithm. In some cases, we can only implement CRNs via the inverse cumulative distribution method, which is computationally demanding. In this situation we can instead fix the number iterations beforehand; the convergence of the algorithm is not crucial, as typically only the initial iterations generate substantial reductions in the variance of the likelihood estimate (\citeNP{dlmrd2012}). For this reason, we recommend a non-strict convergence criterion in Algorithm \ref{alg1}.

Algorithm \ref{alg1} can be subject to numerical instability leading to the divergence of $a_{1:n}$, especially when the state vector $x_t$ is multivariate and when using the natural sampler $p(x_{1:n})$ to draw the initial set of state trajectories. \citeN{KLS3} argue that we can typically eliminate this problem by reducing the step size at the initial iterations of the algorithm. We can achieve this by replacing the measurement density $p(y_{t}|x_{t}^{(s)})$ in \eqref{eq:mincrit} by $p(y_{t}|x_{t}^{(s)})^{\zeta_k}$, where $\zeta_k\in(0,1]$ gradually increases with $k$. Numerical errors may also indicate the use of an excessively low number of samples $S$ to compute the regressions.

\begin{algorithm}
\caption{Efficient importance parameters}
\label{alg1}
\begin{algorithmic}
\vspace{5pt}
\State $\rhd$ Initialise the iteration index $k\gets0$. \vspace{5pt}
\State $\rhd$ Set the initial values for the importance parameters $a_{1:n}^{[0]}$ and denote the associated importance density as
$q^{[0]}(x_{1:n}|y_{1:n})$. A generic and easy to implement choice to initialise the algorithm is the natural sampler, i.e. $q^{[0]}(x_{1:n}|y_{1:n})=p(x_{1:n})$.\vspace{5pt}
\State $\rhd$ Draw a set of common random numbers (CRN) $u_{1:S}$. \vspace{10pt}
\While{convergence criterion is not met}\vspace{15pt}
\State $\rhd$ $k\gets k+1$  \vspace{5pt}
\State \noindent $\rhd$ Obtain $S$ trajectories $x_{1:n}^{(s)}\sim q^{[k-1]}(x_{1:n}|y_{1:n})$ using the CRNs $u_{1:S}$.\vspace{10pt}
\For{t=n:-1:1}\vspace{5pt}
      \State $\rhd$ Solve the least squares problem
      \begin{flalign}\label{eq:eisreg} a_t^{[k]},\gamma_t^{[k]}=\underset{a_t,\gamma_t}{\textrm{argmin}} \sum_{s=1}^{S} \lambda(y_t,x_t^{(s)},x_{t-1}^{(s)},a_t,a_{t+1}^{[k]},\gamma_t)^2\end{flalign}
      \State where
      \begin{equation}\label{eq:mincrit}\lambda(y_t,x_t^{(s)},x_{t-1}^{(s)},a_t,a_{t+1}^{[k]},\gamma_t)= \log\left( \frac{p(y_{t}|x_{t}^{(s)})p(x_{t}^{(s)}|x_{t-1}^{(s)})\chi(x_{t}^{(s)};a^{[k]}_{t+1})}{\gamma_t k(x_t^{(s)},x_{t-1}^{(s)};a_t)}\right),\end{equation}
      \State with $\chi(x_{n}^{(s)};a^{[k]}_{n+1})\equiv 1$. The normalising constant $\gamma_t$ plays no further role
      \State in the method.
      \vspace{5pt}
\EndFor\vspace{15pt}
\EndWhile
\vspace{5pt}
\State $\rhd$ Set the efficient importance density as  $q(x_{1:n}|y_{1:n})=q^{[k]}(x_{1:n}|y_{1:n})$. \vspace{5pt}
 \end{algorithmic}
\end{algorithm}

Even though we have not made any additional assumptions regarding the state space model, the practical applicability of the EIS method relies on the availability of a kernel $k(x_t,x_{t-1};a_t)$ that is able to accurately approximate the numerator in \eqref{ratio} and which leads to a tractable least squares regression within Algorithm 1. The EIS method becomes less interesting when the minimisation problem is nonlinear, in which case the procedure becomes computationally too expensive. That suggests that the EIS method is potentially applicable when the approximating kernel belongs to the exponential family. Existing applications focus on kernels which are conjugate with $p(x_t|x_{t-1})$ or $p(x_t|x_{t-1})\chi(x_{t};a_{t+1})$.\footnote{\citeN{LR2003} and \citeN{ZR2007} originally considered linear Gaussian and inverse Gamma transitions respectively. Nonlinear transitions with additive Gaussian innovations follow easily from the linear case. \citeN{lr2010} consider truncated normal states.} In section \ref{sec:simulation}, we build on the ideas in \citeN{kl2013} to consider a new case in which the state transition has an additive error that follows the Student's $t$ distribution, leading to a conditionally Gaussian setting that is amenable to the use of exponential family kernels.

Finally, we note that more efficient procedures are available when the state transition equation is linear and Gaussian. In this situation the marginal importance density $q(x_t|y_{1:n})$ is available analytically for a Gaussian sampler, enabling numerical and computational gains over the standard algorithm using the results in \citeN{NAIS} and \citeN{KLS3}.

\section{Particle efficient importance sampling}\label{sec:PEIS}

The particle efficient importance sampling method in this section consists of embedding the period $t$ proposal $q(x_{t}|x_{t-1},y_{1:n})$ obtained by the efficient importance sampling method of \citeN{ZR2007} into an auxiliary particle filter algorithm that combines the numerical efficiency of these sequential densities as approximations to $p(y_{t}|x_{t})p(x_{t}|x_{t-1})\chi(x_{t};a_{t+1})$ with resampling steps that ensure that the variance of the target estimate does not grow exponentially with the time series dimension of the problem. Sections \ref{sec:pm} and \ref{sec:newmethod} motivate and describe the new method. Algorithm 2 provides a pseudo code for implementation.

\subsection{Particle methods}\label{sec:pm}

Particle filtering methods recursively obtain a sequence of particles  $\{x_{1:t}^i\}_{i=1}^N$  and associated weights $\{W_{t}^i\}_{i=1}^N$ that approximate the filtering distribution $p(x_{1:t}|y_{1:t})$ at each time period as
\begin{equation*}
\widehat{p}(x_{1:t}|y_{1:t})=\sum_{i=1}^{N}W_t^i\delta_{x_{1:t}^i}(x_{1:t}),
\end{equation*}
where $\delta_{x_{1:t}^i}(x_{1:t})$ denotes the Dirac delta mass located at $x_{1:t}^i$.

The basic particle filter method is based on the sequential importance sampling (SIS) algorithm. Suppose that at the end of period $t-1$ we have a particle system   $\{x_{1:t-1}^i,W_{t-1}^i\}_{i=1}^N$  which approximates the filtering density $p(x_{1:t-1}|y_{1:t-1})$. Upon the arrival of a new observation $y_t$, SIS updates the particle system by propagating the particles $x_{1:t-1}^i$ using the importance distribution
\begin{equation*}
q(x_{t}^{i}|x_{t-1}^{i},y_{1:n})\propto k(x_t^i,x_{t-1}^i;a_t^i)
\end{equation*}
and reweighing each particle trajectory $x_{1:t}^i$ according to
\begin{equation}\label{eq:sisweights}
w_t^i=W_{t-1}^i\,\,\frac{p(y_t|x_t^i)p(x_t^i|x_{t-1}^i)}{q(x_t^i|x_{t-1}^i,y_{1:n})},
\end{equation}
with corresponding normalised weights calculated as
\begin{equation*}
W_t^i=w_t^i/\sum_{i=1}^{N}w_t^i.
\end{equation*}
At each period, we can also estimate the likelihood contribution $p(y_t|y_{1:t-1})$ as
\begin{equation*}
\widehat{p}(y_t|y_{1:t-1})=\sum_{i=1}^{N}w_t^i.
\end{equation*}

It is straightforward to recognise that the efficient high-dimensional importance sampling method of Section \ref{sec:EIS} is a special case of the SIS method in which the proposal density $q(x_{t}^{i}|x_{t-1}^{i},y_{1:n})$ has the kernel $k(x_t^i,x_{t-1}^i;a_t)$ which we construct according to Algorithm 1. In the EIS method, the importance parameters $a_t$  take into account the whole sample information $y_{1:n}$, but do not depend on the particle trajectory $i$. That contrasts with the use of SIS in the particle filter literature, in which $q(x_{t}^{i}|x_{t-1}^{i},y_{1:n})=q(x_{t}^{i}|x_{t-1}^{i},y_{t})$. We refer to this case as online sequential importance sampling. In the online SIS method, we can tailor the importance parameters in the proposal kernel $k(x_t^i,x_{t-1}^i;a_t^i)$  to each inherited particle (indexed by $i$), but do not use the future observations $y_{t+1:n}$ when selecting $a_t^i$.

The second fundamental ingredient of particle methods is resampling, which reduces the impact of the weight degeneracy problem on the performance of the filter in subsequent periods.  It can be shown that as the number of iterations of the SIS method increases, the normalised weights of the particle system become concentrated on fewer particles. Eventually, the weight of a single particle converges to one; see for example \citeN{chopin2004}.  As a result, the variance of estimates obtained using the SIS method grows exponentially in time. Resampling solves this problem by randomly replicating particles from the current population according to their weights, therefore discarding particles with low probability mass.

The standard sequential importance sampling with resampling (SISR) method resamples $N$ particles $\{x_{t}^n\}_{i=1}^N$ with probabilities $\{W_{t}^i\}_{i=1}^N$ and assigns equal weights $W_{t}^i=1/N$ to all particles at the end of each time period. Several unbiased resampling schemes that improve upon multinomial resampling are proposed in the literature; some examples are systematic resampling (\citeNP{kitagawa1996}) and residual resampling (\citeNP{lc1998}). The effective sample size defined as $ESS=1/\sum_{i=1}^{N}(W_t^{i})^2$ is a standard tool for monitoring the degeneracy of particle systems. Since resampling introduces its own source of error by reducing the number of distinct particles at the current period, a straightforward improvement to the basic algorithm is to perform resampling only when he particle weights reach a certain degeneracy threshold.

\subsection{Particle EIS}\label{sec:newmethod}

Since the EIS algorithm is a sequential importance sampler, a SISR version of the method based on the global importance density $q(x_{1:n}|y_{1:n})$ which uses \eqref{eq:sisweights} as resampling weights follows immediately by using the procedure described in Section \ref{sec:pm}. Even though this approach leads to a valid algorithm, we argue that the standard SISR resampling weights are unbalanced and inefficient in this case because the EIS kernel $k(x_t^i,x_{t-1}^i;a_t^i)$ targets $p(y_{t}^i|x_{t}^i)p(x_{t}^i|x_{t-1}^i)\chi(x_{t}^i;a_{t+1})$, which contrasts to $p(y_{t}|x_{t}^i)p(x_{t}^i|x_{t-1}^i)$ for an online SIS kernel.

The critical step in the particle efficient importance sampling method is the introduction of the forward weights
\begin{equation}\label{eq:fwdweight}
w_t^{+i}=W_{t}^i\,\,\chi(x_{t}^i;a_{t+1}),
\end{equation}
leading to the normalised resampling weights
\begin{equation*}
W_t^{+i}=w_t^{+i}/\sum_{i=1}^{N}w_t^{+i}.
\end{equation*}
We now track the degeneracy of the particle system using the forward effective sample size $ESS^+=1/\sum_{i=1}^{N}(W_t^{+i})^2$ .

The justification for the forward weights follows immediately from the construction of the efficient importance sampler. Since
\[w_t^{+i}\propto W_{t-1}^i\,\,\frac{p(y_t|x_t^i)p(x_t^i|x_{t-1}^i)\chi(x_{t}^i;a_{t+1})}{q(x_t^i|x_{t-1}^i,y_{1:n})},\]
the introduction of the integration constant for the next period $\chi(x_{t}^i;a_{t+1})$ matches the importance density $q(x_t^i|x_{t-1}^i,y_{1:n})$
to its target in the minimisation problem \eqref{eq:eisreg}, appropriately balancing the resampling weights.

The use of alternative resampling weights implies that the particle efficient importance sampling method belongs to the class of auxiliary particle filters (APF) introduced by \citeN{ps1999}. The auxiliary particle filter algorithm is designed to improve the efficiency of online particle filters by incorporating period $t$ information when resampling the particles after period $t-1$, anticipating which particles will be in regions of high probability mass after propagation. However, here we use the APF framework just to obtain correct importance weights and likelihood increment estimates when using the forward weights for resampling.

When the forward effective sample size falls below a threshold after period $t-1$, we store the forward weights $\{w_{t-1}^{+i}\}_{i=1}^N$ and resample $N$ particles $\{x_{t-1}^i\}_{i=1}^N$ with probabilities $\{W_{t-1}^{+i}\}_{i=1}^N$ and set $W_{t-1}^i=1/N$ for all the particles. From the APF algorithm, the importance weights after resampling at the end of period $t-1$ and propagating the particles using the importance density $q(x_t^i|x_{t-1}^i,y_{1:n})$ are
\begin{equation*}
w_t^i=W_{t-1}^i\,\frac{p(y_t|x_t^i)p(x_t^i|x_{t-1}^i)}{\chi(x_{t-1}^i;a_{t})q(x_t^i|x_{t-1}^i,y_{1:n})}=
W_{t-1}^i\,\frac{p(y_t|x_t^i)p(x_t^i|x_{t-1}^i)}{k(x_t^i,x_{t-1}^i,a_t)},
\end{equation*}
where $x_{t-1}^i$ are the particles after resampling. When we perform no resampling at the end of the previous iteration, the calculation of the weights and the estimation follows exactly as in the sequential importance sampling algorithm in Section \ref{sec:pm}

\citeN{psgk2012} gives an estimator of the likelihood contribution $p(y_t|y_{t-1})$ based on the auxiliary particle filter.\footnote{This estimator was previously introduced in the working paper by \citeN{pitt2002}} After resampling with weights \eqref{eq:fwdweight}, the estimate is
\begin{equation*}
 \widehat p(y_t|y_{1:t-1})=\left(\sum_{i=1}^{N}w_{t-1}^{+i}\right)\left(\sum_{i=1}^{N}w_t^i\right),
\end{equation*}
where $\{w_{t-1}^{+i}\}_{i=1}^N$ are the forward weights prior to resampling. Proposition 7.4.1 of \citeN{delmoral2004} establishes the unbiasedness of the general auxiliary particle filter estimator, while \citeN{psgk2012} provide an alternative proof of the same result. Algorithm 2 provides the pseudo code for the particle EIS method.

\begin{algorithm}
\caption{Particle Efficient Importance Sampling}
\label{alg2}
\begin{algorithmic}
\vspace{5pt}
\State $\rhd$ Obtain the efficient importance density $q(x_{1:n}|y_{1:n})$ using Algorithm 1 or one of its variations.
\vspace{5pt}
\State At time $t=1$: \vspace{5pt}
      \For{i=1:N}
      \State $\rhd$ Draw $x_1^i\sim q(x_1|y_{1:n})$.
      \State $\rhd$ Compute the importance weight: \begin{flalign*}w_1^i=\frac{p(y_t|x_1^i)p(x_1^i)}{q(x_1^i|y_{1:n})}.\end{flalign*}
      \EndFor
\State $\rhd$ Calculate the estimate of the likelihood contribution as $\widehat p(y_1)=\sum_{i=1}^{N}w_1^i/N$.\vspace{2pt}
\State $\rhd$ Compute the normalised weights $W_1^i=w_1(x_1^i)/\sum_{i=1}^{N}w_1(x_1^i)$, $i=1,\ldots,N$.\vspace{2pt}
\State $\rhd$ Compute the forward weights $w_1^{+i}=W_1^i\cdot\chi(x_{1};a_{2})$, $i=1,\ldots,N$.\vspace{2pt}
\State $\rhd$ Compute the normalised forward weights $W_1^{+i}=w_1^{+i}/\sum_{i=1}^{N}w_1^{+i}$, $i=1,\ldots,N$.\vspace{2pt}
\State $\rhd$ Compute the effective sample size $ESS=1/\sum_{i=1}^{N}(W_1^{+i})^2$.\vspace{10pt}

\State  At time $t\geq2$:\vspace{5pt}
\State $\rhd$ If the effective sample size is below a certain threshold, resample $N$ particles $\{x_{t-1}^i\}_{i=1}^N$ with probabilities $\{W_{t-1}^{+i}\}_{i=1}^N$ and set $W_{t-1}^i=1/N$, for $i=1,\ldots,N$. Store $\{w_{t-1}^{+i}\}_{i=1}^N$. \vspace{5pt}
\For{n=1:N}
      \State $\rhd$ Draw $x_t^i\sim q(x_t^i|y_t,x_{t-1}^i)$.\vspace{5pt}
      \If{resampling}
      \State $\rhd$ Compute the importance weight \begin{flalign*}w_t^i=W_{t-1}^i\times\frac{p(y_t|x_t^i)p(x_t^i|x_{t-1}^i)}{k(x_t^i,x_{t-1}^i,a_t)}.\end{flalign*}
      \Else
       \State $\rhd$ Compute the importance weight \begin{flalign*}w_t^i=W_{t-1}^i\times\frac{p(y_t|x_t^i)p(x_t^i|x_{t-1}^i)}{q(x_t^i|x_{t-1}^i,y_{1:n})}.\end{flalign*}
      \EndIf
\EndFor

\vspace{10pt}
\State (continued on the next page)
\algstore*{alg2}
\end{algorithmic}
\end{algorithm}

\renewcommand{\thealgorithm}{2}

\begin{algorithm}
\caption{(continued)}
\label{alg2}
\begin{algorithmic}
\algrestore*{alg2}
\vspace{10pt}
\State $\rhd$ Calculate the estimate of the likelihood contribution as $\widehat p(y_t|y_{1:t-1})=\sum_{i=1}^{N}w_t^i$ if there is no resampling and as $\widehat p(y_t|y_{1:t-1})=(\sum_{i=1}^{N}w_{t-1}^{+i})(\sum_{i=1}^{N}w_t^i)$ if there is resampling.
\State $\rhd$ Compute the normalised weights $W_1^i=w_t^i/\sum_{i=1}^{N}w_t^i$, $i=1,\ldots,N$.\vspace{2pt}
\State $\rhd$ Compute the forward weights $w_t^{+i}=W_t^i\cdot\chi(x_{t};a_{t+1})$, $i=1,\ldots,N$.\vspace{2pt}
\State $\rhd$ Compute the normalised forward weights $W_t^{+i}=w_t^{+i}/\sum_{i=1}^{N}w_t^{+i}$, $i=1,\ldots,N$.\vspace{2pt}
\State $\rhd$ Compute the effective sample size $ESS^+=1/\sum_{i=1}^{N}(W_t^{+i})^2$.\vspace{10pt}
\end{algorithmic}
\end{algorithm}

\subsection{Discussion}\label{sec:discussion}

We intuitively expect importance sampling methods providing a global approximation to the smoothing density $p(x_{1:n}|y_{1:n})$ to perform better than online sequential importance sampling methods, which restrict the use of sample information $y_{1:n}$. To formalise this idea and shed light on why efficient importance sampling directly addresses the numerical inefficiency of online sequential importance sampling proposals $q(x_{t}|x_{t-1},y_{t})$, we starting by considering the optimal  (but often infeasible) online sequential importance sampler. The conditionally optimal importance distribution for online SIS, in the sense of minimising the variance of the importance weights at each period, is
\begin{equation}
q^*(x_{t}|x_{t-1},y_{t})=p(x_{t}|x_{t-1},y_{t})=\frac{p(y_{t}|x_t)p(x_{t}|x_{t-1})}{p(y_t|x_{t-1})},
\end{equation}
in which case the importance weight is
\begin{equation}\label{eq:optweight}
w_t^{i*}=p(y_t|x_{t-1})=\chi^*(x_{t-1}^i;a_{t}^i).
\end{equation}

It is well known that the conditionally optimal importance density does not guarantee good performance even when used within an SISR algorithm. The reason is transparent from the efficient importance sampling framework of Section \ref{sec:EIS} and equation \eqref{eq:optweight}: the online sequential importance sampling proposal ignores the integration constants $\chi(x_{t}^i;a_{t+1}^i)$, which may have high variance. Hence, the conditionally optimal importance density, if feasible, can still result in inaccurate estimation and frequent resampling if the variance of $p(y_t|x_{t-1})$ is high.

The EIS method of \citeN{ZR2007} directly addresses this numerical inefficiency by incorporating the integration constant $\chi(x_{t};a_{t+1})$ into the variance minimisation problem \eqref{eq:eisreg}. It straightforward to see that the optimal kernel $k^*(x_t,x_{t-1}^i,a_t)\propto p(y_{t}|x_t)p(x_{t}|x_{t-1})\chi(x_{t}^i;a_{t+1})$ which the EIS method approximates leads to importance weights with zero variance. The particle efficient importance sampling method therefore fully combines the numerical efficiency of global importance densities targeting the smoothing distribution with the benefits of resampling.

We make two qualifications. First, global importance sampling comes at the cost of greater difficulty in designing a high-dimensional proposal $q(x_{1:n}|y_{1:n})$ in comparison with devising the low-dimensional sequential proposal densities $q(x_{t}|x_{t-1},y_{t})$ used in online sequential importance sampling. This task will inevitably be highly model specific. Second, we recall that the importance parameters in the online sequential importance kernel $k(x_t^i,x_{t-1}^i,a_t^i)$ can depend on the particle index, which is not the case with efficient importance sampling. Therefore, online SIS can lead to a more accurate approximation to $p(y_{t}|x_t)p(x_{t}|x_{t-1})$ in particular for any given kernel $k(x_t^i,x_{t-1}^i,a_t^i)$ when compared to existing global importance sampling methods. However, we argue that this extra flexibility in the online SIS method is typically of limited practical value, as it is computationally costly to obtain efficient importance parameters for every particle.

\subsection{Antithetic variables}\label{sec:anti}

Antithetic sampling is a variance reduction method based on generating negatively correlated draws from a sampling density. The technique is often instrumental for the success of importance sampling strategies, see for example \citeN{DK00}. In this section we propose a modification of the particle EIS method in order to incorporate the use of antithetic variables. We focus on a particular setting that encompasses our illustrations in Sections \ref{sec:simulation} and \ref{sec:empirical}.

Suppose that we can formulate the dynamics of $x_t$ under the sequential importance density $q(x_{t}|x_{t-1},y_{1:n})$ using the equation
\begin{equation}\label{eq:sisfunc}
x_t=H_t(x_{t-1},\xi_t;a_t),
\end{equation}
where $H_t(.)$ is a nonlinear function and $\xi_t$ is a random variable following a symmetric distribution, which we assume without loss of generality to have mean zero.  Both $H_t$ and the distribution of $\xi_t$ depend on the state transition and the importance parameters.

Starting from a particle system $\{x_{t-1}^i,W_{t-1}^i\}_{i=1}^{N}$, we implement antithetic variables at period $t$ by drawing $N/2$ innovations $\xi_t^1,\ldots,\xi_t^{N/2}$ and propagating the corresponding first half of particles by using equation \eqref{eq:sisfunc} to calculate $x_t^i=H_t(x_{t-1}^i,\xi_t^i;a_t)$ for $i\leq N/2$. We then compute the antithetic draws as $x_t^i=H_t(x_{t-1}^i,-\xi_t^{i-N/2};a_t)$ for $N/2<i\leq N$. When the forward effective sample size reaches the defined degeneracy level at the end of period $t$, we resample only $N/2$ particles $\{x_{t}^j\}_{j=1}^{N/2}$ with probabilities $\{W_{t}^{+i}\}_{i=1}^N$ and duplicate each of then so that  $x_{t}^{N/2+j}=x_{t}^{j}$ for $j\leq N/2$ after resampling. As before, normalised weights after resampling are $W_{t}^i=1/N$ for all $N$ particles. We follow this procedure at every period, where at time $t=1$ we replace the sampling equation in \eqref{eq:sisfunc} by $x_1=H_1(\xi_1;a_1)$.

In the context of particle EIS, antithetic variables have the side effect of amplifying the loss of information when performing resampling because we reduce particle diversity through duplication. Nevertheless, we have found experimentally that this version of the algorithm strongly outperforms the standard version without variance reduction for the models we consider in Sections \ref{sec:simulation} and \ref{sec:empirical}. The reason for the efficiency gain is that resampling takes place infrequently within the particle EIS method. We therefore adopt antithetic sampling throughout the rest of the paper.

\subsection{Particle smoothing}\label{sec:smoothing}

When a full proposal $q(x_{1:n}|y_{1:n})$ is available, it is straightforward to apply importance sampling to estimate functionals based on the full smoothing density  $p(x_{1:n}|y_{1:n})$ in $O(N)$ operations after we calculate the importance weights \eqref{isweight}. See for example \citeN{DK1}. It is therefore natural to also consider the use of particle EIS for smoothing. Estimating smoothing densities becomes computationally more challenging for particle methods due to resampling, which progressively reduces the number of distinct particles in earlier parts of the sample. Particle methods provide an accurate approximation $p(x_{j:n}|y_{1:n})$ only for $j$ relatively close to $n$. While we can expect the particle EIS method to degenerate slower due to infrequent resampling, the standard algorithm will still suffer from this problem for $j\ll n$.

Alternative smoothing schemes that avoid this problem based on forward filtering-backward smoothing recursions and the generalised two-filter formula have been developed in the literature, e.g. \citeN{gdw2004}. These algorithms often have a computational cost which is proportional to $N^2(R+1)$, where $R$ is the number of resampling steps. More sophisticated algorithms with computing time proportional to $N(R+1)$ are now also available, see for example \citeN{fwt2010}. While an investigation of particle smoothing is out of the scope of this paper, we note that when we are able to successfully implement the particle EIS method and directly target the smoothing distribution $p(x_{1:n}|y_{1:n})$ by an importance sampling approximation, we can expect both the number of resampling steps and the number of particles required to reach a certain level of statistical accuracy to be lower than what is the case for standard algorithms, so that the new method can be a useful tool for particle smoothing.

\section{Simulation study}\label{sec:simulation}

This section investigates how particle EIS compares to the EIS method and standard particle filters for likelihood estimation. Section \ref{sec:models} describes the models in the simulation study, Section \ref{sec:details} discusses the alternative methods and implementation details and Section \ref{sec:simresults} provides the comparison methodology and presents the results.

\subsection{Univariate and bivariate stochastic volatility models}\label{sec:models}

We consider two stochastic volatility (SV) models in our simulation study: a univariate two-factor stochastic volatility model with leverage effects and a simple bivariate specification. We assume the following measurement and transition equations for the univariate specification
\begin{align*}
&y_t=\exp([c+x_{1,t}+x_{2,t}]/2)
\eps_t,\qquad t=1,\ldots,n\nonumber\\
&x_{i,t+1}=\phi_1x_{i,t}+\rho_1\sigma_{i,\eta}\eps_t+\sqrt{1-\rho_i^2}\sigma_{i,\eta}\eta_{i,t},\qquad i=1,2
\end{align*}
where the return innovations are i.i.d. and have the standardised Student's $t$ distribution with $\nu$ degrees of freedom and $1>\phi_1>\phi_2>-1$ for stationarity and identification. We implement simulations with two distinct values for the degrees of freedom parameter: $\nu=10$ and $\nu=100$, representing fat tailed and near Gaussian cases respectively. Likewise, we consider two settings for the state disturbances: in the first, they follow the $\Nn(0,1)$ distribution, while in the second they follow the standardised Student's $t$ distribution with 10 degrees of freedom. The parameters for the simulation exercise reflect typical values found by empirical studies: $\phi_1=0.995$, $\sigma_{1,\eta}^2=0.005$, $\rho_1=-0.2$, $\phi_2=0.9$, $\sigma_{2,\eta}^2=0.03$,  and $\rho_2=-0.5$.

The bivariate stochastic volatility model follows the specification originally suggested by \citeN{Harvey94}. The model is
\begin{equation*}
y_t\sim
\mathcal{MVN}\left(\left( \begin{array}{c}
0\\
0\\ \end{array} \right) ,
\left[ \begin{array}{cc}
\sigma_{1,t}^2& \rho_{t}\sigma_{1,t}\sigma_{2,t}\\
\rho_{t}\sigma_{1,t}\sigma_{2,t} & \sigma_{2,t}^2\\ \end{array} \right]\right), \qquad t=1,\ldots,n,
\end{equation*}
\begin{align*}
\sigma_{1,t}^2=\exp(c_1+x_{1,t}),\qquad\sigma_{2,t}^2=\exp(c_2+x_{2,t}),
 \qquad\rho_{t}=\frac{1-\exp(-c_3-x_{3,t})}{1+\exp(-c_3-x_{3,t})},
\end{align*}
where each state follows an AR(1) process,
\begin{align*}
x_{i,t+1}=\phi_ix_{i,t}+\eta_{i,t}, &\qquad\eta_{i,t}\sim \Nn(0,\sigma_{i,\eta}^2),\qquad i=1,2,3.
\end{align*}
The parameters for the simulation study DGP are $c_1=c_2=0$, $\phi_1=\phi_2=0.98$, $\sigma_{1,\eta}=\sigma_{2,\eta}=0.15$, $c_3=1$, $\phi_3=0.99$, $\sigma_{3,\eta}=0.05$.

\subsection{Alternative methods and implementation details}\label{sec:details}

We implement four alternative likelihood estimation methods: the standard EIS method described in Section \ref{sec:EIS} and three particle filter algorithms. The first particle filter algorithm is the bootstrap filter (BF), which corresponds to the sequential importance sampling resampling (SISR) method outlined in Section \ref{sec:pm} with the state transition density as a proposal distribution, so that $q(x_t|x_{t-1},y_t)=p(x_t|x_{t-1})$. The second particle filter method is SISR using a Gaussian proposal which we construct via a second order Taylor expansion of $p(y_t|x_t^i)p(x_t^i|x_{t-1}^i)$ around its mode (conditional on $x_{t-1}^i$). We only consider this method for the univariate SV model with Gaussian state innovations. We label it SISR(2) in the tables. The final particle filter method is a zero order auxiliary particle filter as in \citeN{ps1999}. As in the BF, the proposal is $q(x_t|x_{t-1},y_t)=p(x_t|x_{t-1})$, but the resampling weights become $W_{t-1}^i\,p(y_t|\mu_t(x_{t-1}^i))$, where $\mu_t(x_{t-1}^i)$ is the mean of $x_t$ given $x_{t-1}^i$ according to the state transition. We denote this method by APF(0) in the tables.

The number of samples for Algorithm 1 is $S=50$. The EIS algorithm for the bivariate SV model follows the computationally efficient algorithm of \citeN{KLS3}. We develop the EIS algorithm for the SV model with Student's $t$ innovations in Appendix A. The algorithm follows \citeN{kl2013} and uses a data augmentation scheme that treats the state disturbances as normal-inverse gamma mixtures. We consider two versions of the method. The first only approximates the Gaussian part of the state transition, while the second does importance sampling for both the Gaussian and inverse gamma components. We refer to the two algorithms partial and full EIS respectively. We find that it is important to use the step size reduction modification to Algorithm 1 mentioned in Section \ref{sec:EIS} to ensure that all EIS implementations are free of occasional numerical instability. We also recommend setting the leverage effect coefficients to zero at the initial iterations of the algorithm for the univariate SV model.

We use systematic resampling in all the particle methods. When running the particle filters, we resample when the effective sample size divided by the number of particles falls below 0.5. In the particle efficient importance sampling method, we resample if the forward effective sample size divided by the number of particles is under 0.9, a choice based on experimentation. We use antithetic variables for variance reduction in the EIS and particle EIS methods. We have implemented all methods efficiently using MATLAB mex files. All the reported computing times are based on a computer equipped with an Intel Xeon 3.40 GHz processor with four cores. They do not involve any parallel processing, except in Table \ref{tab:posterior} of Section \ref{sec:empirical}.

\subsection{Likelihood estimation analysis}\label{sec:simresults}

We implement the simulation study as follows. We draw $500$ trajectories of time series dimensions $n=2,500$, $5,000$ and $10,000$ using the three univariate SV and the bivariate SV data generating processes described in Section \ref{sec:models}. For each realisation, we perform twenty independent log-likelihood evaluations at the DGP parameters using particle efficient importance sampling and the alternative methods listed in Section \ref{sec:details}. The number of particles is $N=50$ for all methods. We estimate the variance for each method as
\begin{equation}
\widehat{\textrm{Var}}(\log \widehat L)=\sum_{i=1}^{500}\left(\sum_{j=1}^{20}\frac{(\widehat{\log L_{i,j}}-\overline{\log L_{i}})^2}{19}\right)\bigg/500,
\end{equation}
where $i$ indexes the DGP realisations, $j$ the independent likelihood evaluations, $\widehat{\log L_{i,j}}$ are the corresponding likelihood estimates, and $\overline{\log L_{i}}$ is the sample average for trajectory $i$, so that $\overline{\log L_{i}}=\sum_{j=1}^{20}\widehat{\log L_{i,j}}/20$.

It is essential to take the computing times into account when comparing the likelihood estimation methods, as we can reduce the variance of any estimator by simply increasing the number of particles. In other words, we are interested in the numerical efficiency of each method for a given computational time. We make a distinction the overhead cost per likelihood evaluation, which mainly corresponds to the time to run Algorithm 1, and the rest of EIS and particle EIS algorithms, for which the computational cost is proportional to the number of particles $N$. We define the efficiency relative to the standard EIS method benchmark as
\begin{equation}\label{eq:efficiency}
\textrm{Efficiency}^{h,N}=\frac{\widehat{\textrm{Var}}(\log \widehat L^{h,N})}{\widehat{\textrm{Var}}(\log \widehat L^{b,N})}\left(1+\frac{\tau_1^b+N\tau_2^b-\tau_1^h-N\tau_2^h}{N\tau_2^h}\right)^{-1},
\end{equation}
where $h$ indexes the method, $b$ indexes the benchmark and $\widehat{\textrm{Var}}(\log \widehat L^{h,N})$ denotes the estimated variance of method $h$ with $N$ particles. We assume that the computing time is an affine function of the number of particles
\begin{equation}\label{eq:ct}
\textrm{Computing time}^{h,N}=\tau_1^h+N\tau_2^h.
\end{equation}
We have that $\tau_1^h=0$ for the particle filters. In the tables we label $\tau_1^h$ and $N\tau_2^h$ as EIS density time and likelihood time respectively. The measurements take into account the resampling steps. Assuming that the variance of the log of the likelihood estimate we obtain using each method scales at rate $1/N$, the efficiency measure estimates the variance associated with algorithm $h$ for a number of particles $N^\prime$ such that $\tau_1^h+N^\prime\tau_2^h=\tau_1^b+N\tau_2^b$. It therefore estimates the variance of the method $h$ estimate when we give it the same total computing time as the benchmark.

Tables \ref{tab:example2}-\ref{tab:example1} present the results. Three main findings appear in all the cases we considered. First, the particle EIS method brings large reductions in variance over the standard EIS method. When $n=10,000$ the decrease in variance ranges from $80\%$ for the univariate SV model with Student's $t$ state disturbances to 95\% for the bivariate specification. These gains come with almost no increase in computational time since the new method resamples infrequently. Second, the use of a global approximation in the EIS and particle EIS methods leads to substantial gains in efficiency over the particle filters. The simulations reveal that even after taking the larger computing times into account,  the particle EIS method is 112 more efficient than the bootstrap filter for the model in Table \ref{tab:example4}, going up to 5,812 times more efficient in the setting of Table \ref{tab:example2}. In contrast, the use of a better importance density for particle filtering in the SISR(2) method is counterproductive when taking into account the excessive computational burden of constructing proposals and computing importance weights for each particle separately. Finally, as expected theoretically, the relative performance of the EIS method deteriorates quickly with the time series dimension, despite its good behaviour in the examples. The particle EIS method completely avoids this problem, approximately maintaining a constant relative performance compared to the particle filters for all time series dimensions.

\begin{table}[!tbp]
\vspace{-20pt}
\begin{center}
{\small \caption{{\sc Two-factor stochastic volatility with leverage effects and Student's $t$ return innovations ($\nu=10$): likelihood evaluation}.}\label{tab:example2}
\vspace{1pt}
\begin{threeparttable}
{\footnotesize The table compares the efficiency of different likelihood estimation methods. The methods are the bootstrap filter (BF), sequential importance sampling with resampling based on a Laplace approximation (SISR (2)), a zero order auxiliary particle filter (APF), efficient importance sampling (EIS) and particle EIS (P-EIS). }
\vspace{7pt}
\begin{tabular}{lccccc}
\hline\hline
& \multicolumn{ 5}{c}{$n=2,500$} \\
 & BF & SISR (2) & APF (0) & EIS & P-EIS \\
 \cline{2-6}
Variance & 8.84 & 7.45 & 8.41 & 0.002 & 0.001 \\
Variance ratio & 4498 & 3790 & 4282 & 1.000 & 0.462 \\
EIS density time & - & - & - & 0.391 & 0.391 \\
Likelihood time & 0.030 & 0.761 & 0.026 & 0.048 & 0.049 \\
Efficiency ($N=50$) & 312 & 6582 & 254 & 1.000 & 0.477 \\
Efficiency ($N\rightarrow\infty$)  & 2874 & 60642 & 2336 & 1.000 & 0.477 \\
\\
 & \multicolumn{ 5}{c}{$n=5,000$} \\
 & BF & SISR (2) & APF (0) & EIS & P-EIS \\
\cline{2-6}
Variance & 17.45 & 14.41 & 17.59 & 0.007 & 0.002 \\
Variance ratio & 2394 & 1976 & 2412 & 1.000 & 0.257 \\
EIS density time & - & - & - & 0.759 & 0.759 \\
Likelihood time & 0.059 & 1.496 & 0.052 & 0.090 & 0.097 \\
Efficiency ($N=50$) & 167 & 3483 & 147 & 1.000 & 0.279 \\
Efficiency ($N\rightarrow\infty$)  & 1581 & 33016 & 1392 & 1.000 & 0.279 \\
\\
 & \multicolumn{ 5}{c}{$n=10,000$} \\
 & BF & SISR (2) & APF (0) & EIS & P-EIS \\
\cline{2-6}
Variance & 34.92 & 29.19 & 34.06 & 0.027 & 0.004 \\
Variance ratio & 1273 & 1064 & 1241 & 1.000 & 0.141 \\
EIS density time & - & - & - & 1.384 & 1.384 \\
Likelihood time & 0.109 & 2.917 & 0.100 & 0.155 & 0.170 \\
Efficiency ($N=50$) & 90 & 2016 & 81 & 1.000 & 0.154 \\
Efficiency ($N\rightarrow\infty$)  & 894 & 19995 & 803 & 1.000 & 0.154 \\
\hline\hline
\end{tabular}
\end{threeparttable}}
\end{center}
\end{table}

\begin{table}[!tbp]
\vspace{-20pt}
\begin{center}
{\small \caption{{\sc Two-factor stochastic volatility with leverage effects and Student's $t$ return innovations ($\nu=100$): likelihood evaluation}.}\label{tab:example3}
\vspace{1pt}
\begin{threeparttable}
{\footnotesize The table compares the efficiency of different likelihood estimation methods for the two-factor stochastic volatility model. The methods are the bootstrap filter (BF), sequential importance sampling with resampling based on a Laplace approximation (SISR (2)), a zero order auxiliary particle filter (APF), efficient importance sampling (EIS) and particle EIS (P-EIS). }
\vspace{7pt}
\begin{tabular}{lccccc}
\hline\hline
 & \multicolumn{ 5}{c}{$n=2,500$} \\
 & BF  & SISR (2) & APF (0) & EIS & P-EIS \\
\cline{2-6}
Variance & 10.72 & 8.65 & 10.34 & 0.011 & 0.003 \\
Variance ratio & 978 & 788 & 943 & 1.000 & 0.245 \\
EIS density time & - & - & - & 0.352 & 0.352 \\
Likelihood time & 0.030 & 0.758 & 0.027 & 0.048 & 0.047 \\
Efficiency ($N=50$) & 73 & 1494 & 64 & 1.000 & 0.239 \\
Efficiency ($N\rightarrow\infty$)  & 602 & 12390 & 534 & 1.000 & 0.239 \\
 &  &  &  &  &  \\
 & \multicolumn{ 5}{c}{$n=5,000$} \\
 & BF  & SISR (2) & APF (0) & EIS & P-EIS \\
\cline{2-6}
Variance & 20.98 & 16.86 & 20.32 & 0.039 & 0.005 \\
Variance ratio & 542 & 436 & 525 & 1.000 & 0.136 \\
EIS density time & - & - & - & 0.699 & 0.699 \\
Likelihood time & 0.057 & 1.500 & 0.055 & 0.092 & 0.093 \\
Efficiency ($N=50$) & 39 & 826 & 36 & 1.000 & 0.137 \\
Efficiency ($N\rightarrow\infty$)  & 338 & 7104 & 313 & 1.000 & 0.137 \\
 &  &  &  &  &  \\
 & \multicolumn{ 5}{c}{$n=10,000$} \\
 & BF & SISR (2) & APF (0) & EIS & P-EIS \\
\cline{2-6}
Variance & 40.59 & 33.41 & 41.17 & 0.152 & 0.011 \\
Variance ratio & 268 & 220 & 272 & 1.000 & 0.070 \\
EIS density time & - & - & - & 1.404 & 1.404 \\
Likelihood time & 0.106 & 2.945 & 0.105 & 0.158 & 0.158 \\
Efficiency ($N=50$) & 18 & 415 & 18 & 1.000 & 0.070 \\
Efficiency ($N\rightarrow\infty$)  & 180 & 4099 & 181 & 1.000 & 0.070 \\
\hline\hline
\end{tabular}
\end{threeparttable}}
\end{center}
\end{table}

\begin{table}[!tbp]
\vspace{-25pt}
\begin{center}
{\small \caption{{\sc Two-factor stochastic volatility with leverage effects and Student's $t$ return and state innovations: likelihood evaluation}.}\label{tab:example4}
\vspace{1pt}
\begin{threeparttable}
{\footnotesize The table compares the efficiency of different likelihood estimation methods for the two-factor stochastic volatility model with Student's $t$ state disturbances. The methods are the bootstrap filter (BF), a zero order auxiliary particle filter (APF), efficient importance sampling (EIS) and particle EIS (P-EIS). The EIS methods are based on a data augmentation scheme for the transition density. The full EIS method performs importance sampling in both the Gaussian and the inverse-gamma state components, whereas the partial EIS method performs importance sampling only on the Gaussian component(see Appendix A for the details). }
\vspace{7pt}
\begin{tabular}{lcccccc}
\hline\hline
 & \multicolumn{ 6}{c}{$n=2,500$} \\
 & & &\multicolumn{2}{c}{EIS} & \multicolumn{2}{c}{P-EIS} \\
 & BF & APF (0) & partial & full & partial & full \\
\cline{2-7}
Variance & 8.77 & 8.77 & 0.075 & 0.031 & 0.040 & 0.016 \\
Variance ratio & 117 & 117 & 1.000 & 0.410 & 0.541 & 0.219 \\
EIS density time & - & - & 1.245 & 1.258 & 1.245 & 1.258 \\
Likelihood time & 0.031 & 0.031 & 0.183 & 0.181 & 0.180 & 0.184 \\
Efficiency ($N=50$) & 2.517 & 2.579 & 1.000 & 0.406 & 0.534 & 0.221 \\
Efficiency ($N\rightarrow\infty$)  & 19.684 & 20.169 & 1.000 & 0.406 & 0.534 & 0.221 \\
\\
 & \multicolumn{ 6}{c}{$n=5,000$} \\
 & & &\multicolumn{2}{c}{EIS} & \multicolumn{2}{c}{P-EIS} \\
 & BF & APF (0) & partial & full & partial & full \\
\cline{2-7}
Variance & 17.45 & 17.38 & 0.217 & 0.089 & 0.079 & 0.030 \\
Variance ratio & 80 & 80 & 1.000 & 0.410 & 0.365 & 0.137 \\
EIS density time & - & - & 2.184 & 2.249 & 2.184 & 2.249 \\
Likelihood time & 0.056 & 0.057 & 0.303 & 0.319 & 0.331 & 0.338 \\
Efficiency ($N=50$) & 1.801 & 1.837 & 1.000 & 0.432 & 0.399 & 0.153 \\
Efficiency ($N\rightarrow\infty$)  & 14.775 & 15.070 & 1.000 & 0.432 & 0.399 & 0.153 \\
\\
& \multicolumn{ 6}{c}{$n=10,000$} \\
 & & &\multicolumn{2}{c}{EIS} & \multicolumn{2}{c}{P-EIS} \\
 & BF & APF (0) & partial & full & partial & full \\
 \cline{2-7}
Variance & 35.96 & 34.20 & 0.644 & 0.291 & 0.155 & 0.056 \\
Variance ratio & 56 & 53 & 1.000 & 0.453 & 0.240 & 0.087 \\
EIS density time & - & - & 3.556 & 3.605 & 3.556 & 3.605 \\
Likelihood time & 0.097 & 0.102 & 0.511 & 0.529 & 0.534 & 0.556 \\
Efficiency ($N=50$) & 1.334 & 1.337 & 1.000 & 0.468 & 0.251 & 0.095 \\
Efficiency ($N\rightarrow\infty$)  & 10.622 & 10.645 & 1.000 & 0.468 & 0.251 & 0.095 \\
\hline\hline
\end{tabular}
\end{threeparttable}}
\end{center}
\end{table}

\begin{table}[!tbp]
\vspace{-20pt}
\begin{center}
{\small \caption{{\sc Bivariate stochastic volatility: likelihood evaluation}.}\label{tab:example1}
\vspace{1pt}
\begin{threeparttable}
{\footnotesize The table compares the efficiency of different likelihood estimation methods for the bivariate stochastic volatility model. The methods are the bootstrap filter (BF), a zero order auxiliary particle filter (APF), efficient importance sampling (EIS) and particle EIS (P-EIS). }
\vspace{7pt}
\begin{tabular}{lcccc}
\hline\hline
 & \multicolumn{ 4}{c}{$n=2,500$} \\
 & BF  & APF (0) & EIS & P-EIS \\
\cline{2-5}
Variance & 48.86 & 45.25 & 0.087 & 0.012 \\
Variance ratio & 564 & 522 & 1.000 & 0.138 \\
EIS density time & - & - & 0.146 & 0.146 \\
Likelihood time & 0.027 & 0.028 & 0.032 & 0.032 \\
Efficiency ($N=50$) & 86 & 82 & 1.000 & 0.141 \\
Efficiency ($N\rightarrow\infty$)  & 481 & 461 & 1.000 & 0.141 \\
\\
 & \multicolumn{ 4}{c}{$n=5,000$} \\
 & BF  & APF (0) & EIS & P-EIS \\
\cline{2-5}
Variance & 97.41 & 91.47 & 0.313 & 0.023 \\
Variance ratio & 311 & 292 & 1.000 & 0.075 \\
EIS density time & - & - & 0.385 & 0.385 \\
Likelihood time & 0.056 & 0.058 & 0.073 & 0.072 \\
Efficiency ($N=50$) & 38 & 37 & 1.000 & 0.074 \\
Efficiency ($N\rightarrow\infty$)  & 239 & 231 & 1.000 & 0.074 \\
\\
 & \multicolumn{ 4}{c}{$n=10,000$} \\
 & BF & APF (0) & EIS & P-EIS \\
\cline{2-5}
Variance & 192 & 181 & 1.003 & 0.048 \\
Variance ratio & 191 & 180 & 1.000 & 0.048 \\
EIS density time & - & - & 0.833 & 0.833 \\
Likelihood time & 0.108 & 0.111 & 0.152 & 0.149 \\
Efficiency ($N=50$) & 21 & 20 & 1.000 & 0.047 \\
Efficiency ($N\rightarrow\infty$)  & 136 & 132 & 1.000 & 0.047 \\
\hline\hline
\end{tabular}
\end{threeparttable}}
\end{center}
\end{table}

\section{Empirical application}\label{sec:empirical}

This section studies the performance of the particle EIS method as a tool for Bayesian inference. We consider an empirical application of the bivariate stochastic volatility model of Section \ref{sec:models} using daily holding period returns for IBM and General Electric stocks between 1990 and 2012. The total number of bivariate time series observations is 5,797. The source of the series is the Center for Research in Security Prices (CRSP) database.  We adopt the following independent priors for each parameter
\begin{eqnarray*}
&c_i\sim \Nn(0,1), \qquad \phi_i\sim \textrm{Unif}(0,1),\,\,\,\,i=1,2,3,\nonumber\\
&\sigma_{i,\eta}^2\sim \textrm{IG}(2.5,0.035),\,\,\,\,i=1,2, \qquad \sigma_{3,\eta}^2\sim \textrm{IG}(2.5,0.0075), \nonumber
\end{eqnarray*}
where $\textrm{IG}(a,b)$ denotes the inverse Gamma distribution with shape $a$ and scale $b$.

We investigate two approaches for posterior inference: particle marginal Metropolis-Hastings (PMMH, \citeNP{pmcmc}) and importance sampling squared (IS$^2$, \citeNP{tspk2013}). The key idea of both PMMH and IS$^2$ is that replacing the unknown true likelihood by an unbiased estimator in standard Metropolis-Hastings and IS algorithms still leads to valid procedures that target the correct posterior distribution of the parameters. Let $p(\theta)$ be the prior distribution, $p(y_{1:n}|\theta)$ the likelihood \eqref{lik} and $\pi(\theta)\propto p(y_{1:n}|\theta)p(\theta)$ the posterior distribution of the parameters defined on $\Theta$. Suppose we want to calculate the integral

\[\pi(\varphi)=\int_\Theta \varphi(\theta)\pi(\theta) \dd \theta.\]

The IS$^2$ method involves the following steps

\begin{enumerate}
\item Draw $M$ parameter samples $\theta_1,\ldots,\theta_M$ from an importance density $q(\theta|y_{1:n})$.

\item Compute an unbiased estimate $\widehat p(y_{1:n}|\theta_i)$ of the likelihood function for $i=1,\ldots,M$.

\item Compute the importance weights for $i=1,\ldots,M$
\[\omega(\theta_i,y)=\frac{\widehat p(y_{1:n}|\theta_i)p(\theta_i)}{q(\theta_i|y_{1:n})}\]

\item Compute the importance sampling estimator

\[\widehat\pi(\varphi)=\frac{\sum_{i=1}^M \varphi(\theta_i)\omega(\theta_i,y_{1:n})}{\sum_{i=1}^M \omega(\theta_i,y_{1:n})}.\]

\item We can also estimate the marginal likelihood $p(y_{1:n})=\int_\Theta p(y_{1:n}|\theta)p(\theta) \dd \theta$ as $\widehat p(y_{1:n})=\sum_{i=1}^M \omega(\theta_i,y_{1:n})/M$.
\end{enumerate}

To obtain the parameter proposals $q(\theta|y_{1:n})$ for the $IS^2$ and particle independent Metropolis-Hastings (PIMH) methods, we consider the mixture of $t$ by importance sampling weighted expectation maximisation (MitISEM) method of \citeN{hod2012}. The MitISEM method implements a recursive sequence of importance weighted expectation maximisation that minimises the Kullback-–Leibler divergence between the posterior distribution and a mixture of Student's $t$ densities proposal.

We implement the basic proposal training algorithm in that paper, but replace the true likelihood used in the original method by estimates provided by the EIS and particle EIS methods with $N=50$ particles. We label these two cases MitISEM (EIS) and MitISEM (P-EIS) respectively. We use 250 points from a Halton sequence with 9 dimensions and 250 antithetic draws to generate samples from the candidate densities within the training phase of the algorithm.  We found that a multivariate Student's $t$ density provides a good approximation to the posterior for the current problem. In our illustrations, the likelihood estimation algorithm which we use when running the IS$^2$ and PIMH algorithms does not necessarily correspond to the one we adopt for training the MitISEM proposal. Our objective in doing so is to study the performance of different unbiased likelihood estimation methods when the proposal is fixed.

\subsection{Choosing the number of particles}\label{sec:optN}

\citeN{psgk2012} and \citeN{tspk2013} study efficient implementations of Markov chain Monte Carlo and importance sampling when using unbiased likelihood estimators and general parameter proposals. The idea behind these papers is that the choice of the number of particles for likelihood estimation is a trade-off between variance reduction and computing time, which we may best allocate running more iterations of the Markov chain or generating additional importance samples for the parameters.

Assume that the log of the likelihood estimator is normal and that its variance is constant across different values of $\theta$. The main finding in these papers is that the optimal number of particles to minimise the computing time for any given target Monte Carlo variance is such that the variance of the log-likelihood estimator is approximately equal to one when using particle filters. The EIS and particle EIS methods involve the additional complication of the overhead associated with Algorithm 1, which does not depend on $N$. Let the variance of the log-likelihood estimator be $\textrm{Var}(\log \widehat L^h)/N$. The optimal number of particles is
\[
N_h^{\textrm{opt}}=\frac{\textrm{Var}(\log \widehat L^{h})\(1+\sqrt{1+4\textrm{Var}(\log \widehat L^h)^{-1}(\tau_1^h/\tau_2^h)}\)}{2},
\]
where $\tau_1^h$ and $\tau_2^h$ are defined in \eqref{eq:ct}. Note that $N_h^{\textrm{opt}}>\textrm{Var}(\log \widehat L^h)$ when $\tau_1^h>0$. By dividing the variance of the log-likelihood by $N_h^{\textrm{opt}}$, we can see that the optimal variance of the log-likelihood estimate is lower than one when there is an overhead cost for estimating the likelihood.

Table \ref{tab:example5} summarises a limited simulation study of how the variance of the log of the estimated likelihood depends on the method. The motivation for the study is to determine the number of particles for the empirical example. We carry out the simulation study as follows.  First, we obtain a proposal density that approximates the posterior distribution of the parameters using the MitISEM (EIS) method. We then generate $M=100$ draws from this proposal. For each sampled parameter vector, we perform 20 independent log-likelihood evaluations using the bootstrap filter, the EIS and the particle EIS algorithms.  We use $S=32$ simulations to obtain the importance parameters in the EIS method. We report the average of the sample variances across the 100 parameters draws, the corresponding variance ratios (with the EIS method as the benchmark), the computing time for obtaining the efficient importance density ($\tau_1^h$), the computing time for the likelihood estimation step ($N\tau_2^h$), and the relative efficiency as defined in \eqref{eq:efficiency}.

Consistent with Table \ref{tab:example1}, we find a 96.5\% reduction in average variance for particle EIS in comparison with the EIS method. The results imply that the optimal number of particles is approximately $14,800$ for the bootstrap filter, $310$ for EIS, and $42$ for particle EIS. That leads us to use $N=150$ and $N=300$ samples for the EIS method and $N=10$ and $N=50$ particles for P-EIS, with the lower number of particles indicating the case for which the variance of the log-likelihood estimate is approximately one on average. For the bootstrap filter, we set the number of particles sub-optimally to $N=5,000$ due to the excessively high computational cost of an ideal implementation for this problem.

The theoretical results on the optimal implementation of PMMH and $IS^2$, in conjunction with Tables \ref{tab:example2}-\ref{tab:example4}, highlight that the EIS method is remarkably efficient for Bayesian inference in the univariate SV model with Student's $t$ return innovations. Based on the variance estimates for the EIS method in those tables, the standard algorithm with no resampling requires only 2 to 16 samples (including antithetic draws) to achieve a log-likelihood variance of approximately one for $n$ as large as $10,000$. For particle EIS, only two particles are typically sufficient in this scenario. For this reason, we focus on the more challenging bivariate specification in this section.

\begin{table}[!tbp]
\vspace{-20pt}
\begin{center}
{\small \caption{{\sc Bivariate SV - likelihood evaluation for the parameters sampled from the multivariate $t$ proposal}.}\label{tab:example5}
\vspace{1pt}
\begin{threeparttable}
{\footnotesize }
\vspace{7pt}
 \begin{tabular}{lccc}
\hline\hline
 & BF (N=100) & EIS (N=100) & P-EIS (N=100) \\
\cline{2-4}
Variance & 14.786 & 1.582 & 0.055 \\
Variance ratio & 9.349 & 1.000 & 0.035 \\
EIS density time ($\tau_1^h$) & - & 0.567 & 0.567 \\
Likelihood time ($N\tau_2^h$) & 1.039 & 0.188 & 0.205 \\
Efficiency & 12.858 & 1.000 & 0.038 \\
\hline\hline
\end{tabular}
\end{threeparttable}}
\end{center}
\end{table}

\subsection{Posterior analysis}

\begin{figure}
    \begin{center}
        \subfigure[$c_1$]{%
           \includegraphics[scale=0.35]{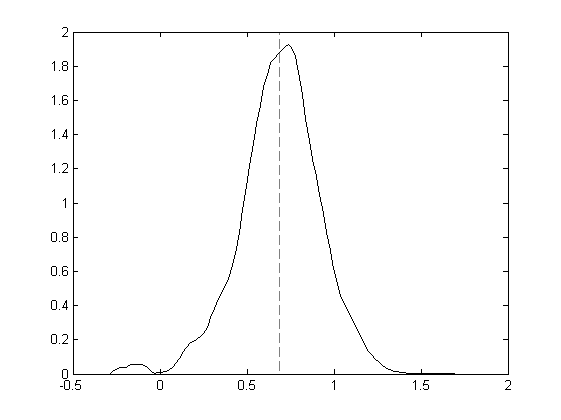}
        }%
        \subfigure[$\phi_1$]{%
           \includegraphics[scale=0.35]{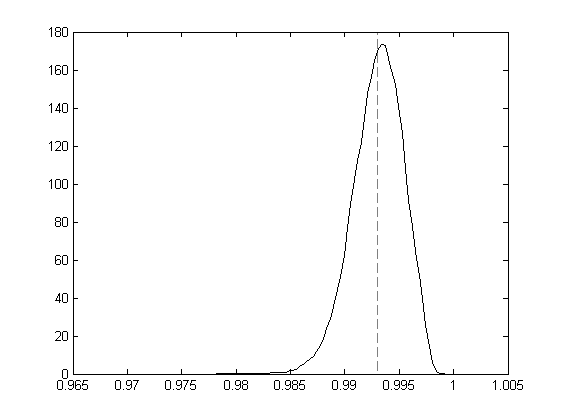}
        }%
        \subfigure[$\sigma_1^2$]{%
           \includegraphics[scale=0.35]{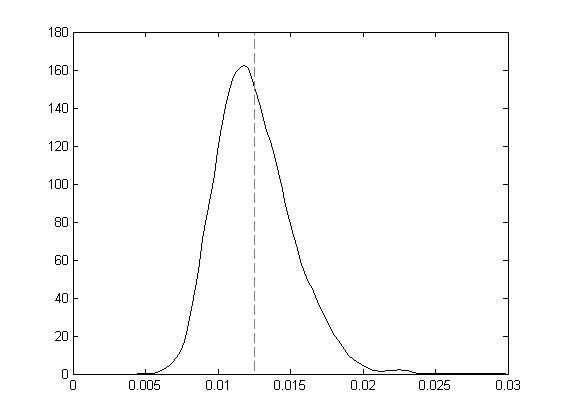}
        }%
        \\
        \subfigure[$c_2$]{%
           \includegraphics[scale=0.35]{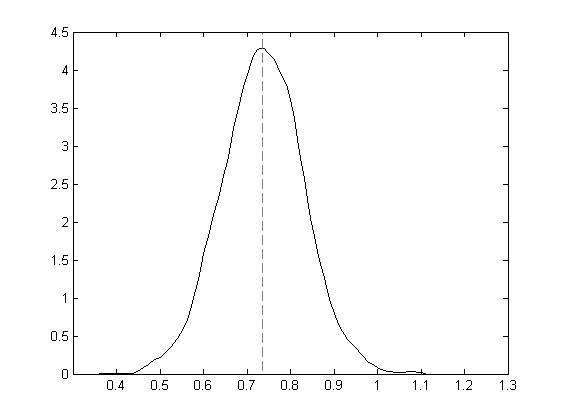}
        }%
        \subfigure[$\phi_2$]{%
           \includegraphics[scale=0.35]{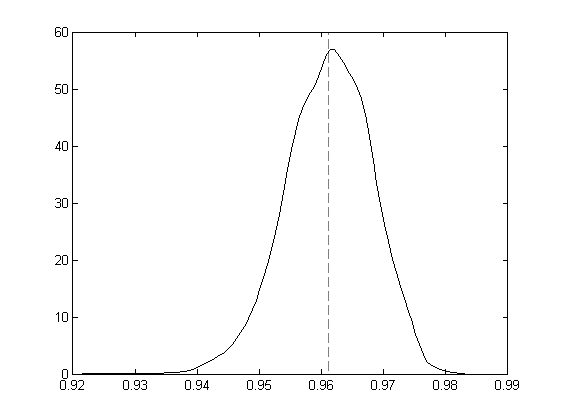}
        }%
        \subfigure[$\sigma_2^2$]{%
           \includegraphics[scale=0.35]{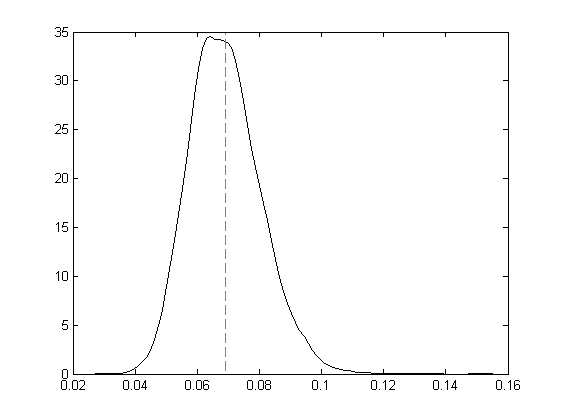}
        }%
        \\
        \subfigure[$c_3$]{%
           \includegraphics[scale=0.35]{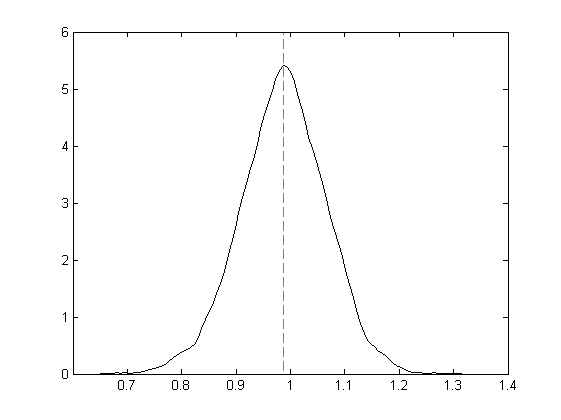}
        }%
        \subfigure[$\phi_3$]{%
           \includegraphics[scale=0.35]{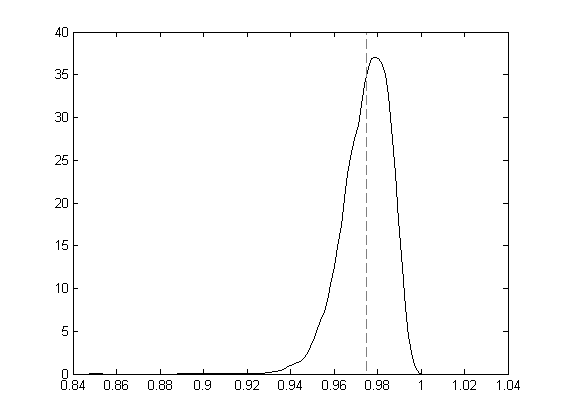}
        }%
        \subfigure[$\sigma_3^2$]{%
           \includegraphics[scale=0.35]{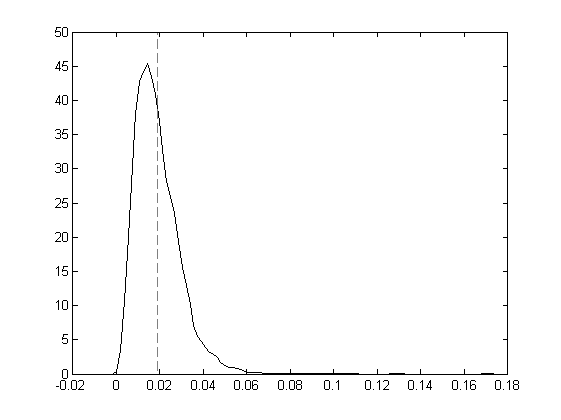}
        }%
    \end{center}
    \caption{Kernel smoothing density estimates of the marginal posterior distributions of the bivariate SV model parameters (estimated by importance sampling squared). The dashed lines indicate the posterior means. }\label{fig:posterior}
\end{figure}

Table \ref{tab:posterior} presents estimates of selected posterior distribution statistics estimated by the IS$^2$ method. We estimate the likelihood for a given set of parameters using the particle EIS method with $N=50$ particles. We estimated the posterior distribution using $M=10,000$ importance samples for the parameters, which required a total computing time of 21 minutes (parallelising the computations over 4 cores).  We also estimate the Monte Carlo standard errors by bootstrapping the importance samples. The low MC standard errors confirm the efficiency of IS$^2$ approach using particle EIS. Figure \ref{fig:posterior} estimates the kernel smoothing density estimates of the marginal posteriors.

\begin{table}[!tbp]
\vspace{-20pt}
\begin{center}
{\small \caption{{\sc Bivariate SV - posterior statistics estimated by importance sampling squared}.}\label{tab:posterior}
\vspace{1pt}
\begin{threeparttable}
{\footnotesize The table presents estimates of selected posterior distribution statistics for the bivariate stochastic volatility application. The Monte Carlo standard errors are in brackets.}
\vspace{7pt}
\begin{tabular}{lcccccc}
\hline\hline
& Mean & Std. Dev. & Skew. & Kurt. & \multicolumn{2}{c}{90\% Credible Interval} \\
\cline{2-7}
$c_1$ & $\underset{[0.014]}{0.687}$ & $\underset{[0.008]}{0.229}$ & $\underset{[0.051]}{-0.469}$ & $\underset{[0.131]}{4.188}$ & $\underset{[0.044]}{0.299}$ & $\underset{[0.007]}{1.048}$ \\[10pt]
$\phi_1$ & $\underset{[<0.001]}{0.993}$ & $\underset{[<0.001]}{0.002}$ & $\underset{[0.039]}{-0.465}$ & $\underset{[0.070]}{3.237}$ & $\underset{[<0.001]}{0.989}$ & $\underset{[<0.001]}{0.997}$ \\[10pt]
$\sigma_1^2$ & $\underset{[<0.001]}{0.013}$ & $\underset{[<0.001]}{0.003}$ & $\underset{[0.046]}{0.647}$ & $\underset{[0.155]}{3.753}$ & $\underset{[<0.001]}{0.009}$ & $\underset{[<0.001]}{0.017}$ \\[10pt]
$c_2$ & $\underset{[0.002]}{0.736}$ & $\underset{[0.001]}{0.093}$ & $\underset{[0.059]}{0.008}$ & $\underset{[0.087]}{3.152}$ & $\underset{[0.003]}{0.585}$ & $\underset{[0.005]}{0.886}$ \\[10pt]
$\phi_2$ & $\underset{[<0.001]}{0.961}$ & $\underset{[<0.001]}{0.007}$ & $\underset{[0.047]}{-0.288}$ & $\underset{[0.049]}{3.079}$ & $\underset{[<0.001]}{0.950}$ & $\underset{[<0.001]}{0.972}$ \\[10pt]
$\sigma_2^2$ & $\underset{[<0.001]}{0.069}$ & $\underset{[<0.001]}{0.011}$ & $\underset{[0.034]}{0.415}$ & $\underset{[0.079]}{3.246}$ & $\underset{[<0.001]}{0.052}$ & $\underset{[<0.001]}{0.089}$ \\[10pt]
$c_3$ & $\underset{[0.001]}{0.987}$ & $\underset{[0.001]}{0.078}$ & $\underset{[0.060]}{-0.124}$ & $\underset{[0.159]}{3.270}$ & $\underset{[0.004]}{0.855}$ & $\underset{[0.002]}{1.110}$ \\[10pt]
$\phi_3$ & $\underset{[<0.001]}{0.975}$ & $\underset{[<0.001]}{0.011}$ & $\underset{[0.031]}{-0.764}$ & $\underset{[0.084]}{3.866}$ & $\underset{[<0.001]}{0.955}$ & $\underset{[0.001]}{0.990}$ \\[10pt]
$\sigma_3^2$ & $\underset{[<0.001]}{0.019}$ & $\underset{[<0.001]}{0.010}$ & $\underset{[0.036]}{1.145}$ & $\underset{[0.137]}{5.277}$ & $\underset{[<0.001]}{0.006}$ & $\underset{[<0.001]}{0.037}$ \\
\hline\hline
\end{tabular}
\end{threeparttable}}
\end{center}
\end{table}

\subsection{Importance sampling squared}

This section compares the use of the bootstrap filter, EIS and P-EIS algorithms for the IS$^2$ method.  We consider the efficiency of each method for estimating the posterior mean of all the parameters and the marginal likelihood. We estimate the Monte Carlo variances associated with each method as in Section \ref{sec:simresults} by running 250 independent replications of the importance sampling algorithm using $M=500$ importance samples for the parameters. We consider two versions of the MitISEM method: one using the EIS method for estimating the likelihood the training step of the method
the importance density and another using the P-EIS method for the same purpose. Our efficiency measure is the time normalised variance of estimates, which we define as the product of Monte Carlo variance and the computational time.  We report all the time normalised variances as relative to the EIS method with $N=150$ samples. The time normalised variance determines the total computing time required for obtaining any given MC variance for the posterior mean and marginal likelihood using each method.

Table \ref{tab:example7} summarises the results. Focusing on the MitISEM (EIS) proposal, the table shows reductions in time normalised variance which range from 79\% to 97\% for the P-EIS method relative to the EIS method. When considering the P-EIS method for constructing the proposal, we find further reductions in time normalised variance of as much as 60\%. Compared to the bootstrap filter, the reductions in time normalised variance range from 99.09\% to 99.81\%. The table also shows that the EIS method with $N=300$ samples, which should be approximately the optimal number of samples according to Section \ref{sec:optN}, has an inferior performance to the implementation with $N=150$. We conjecture that this is because the EIS log-likelihood estimates are skewed for $N=150$ and $N=300$, whereas the theoretical result for the optimal number of particles is based on a normality assumption. In this setting, increasing the number of samples from 150 to 300 reduces the EIS log-likelihood variance by less than 50\%, so that the benefit of increasing $N$ is lower than assumed by the result on the optimal number of samples.

\begin{table}[!tbp]
\vspace{-20pt}
\begin{center}
{\small \caption{{\sc Bivariate SV - relative time normalised variances for posterior inference using IS$^2$}.}\label{tab:example7}
\vspace{1pt}
\begin{threeparttable}
{\footnotesize The table shows the performance of different methods for estimating the posterior distribution of the bivariate stochastic volatility model by IS$^2$. We indicate the number of particles for each method by $N$. We also report the average computing time in seconds. }
\vspace{7pt}
\begin{tabular}{lcccccccc}
\hline\hline
 & \multicolumn{ 5}{c}{MitISEM  (EIS)} &  & \multicolumn{ 2}{c}{MitISEM (PEIS)} \\
 & BF & \multicolumn{ 2}{c}{EIS} & \multicolumn{ 2}{c}{P-EIS} &  & \multicolumn{ 2}{c}{P-EIS} \\
 & N=5,000 & N=150 & N=300 & N=10 & N=50 &  & N=10 & N=50 \\
\cline{2-6}\cline{8-9}
$c_1$ & 14.951 & 1.000 & 1.824 & 0.359 & 0.210 &  & 0.122 & 0.136 \\
$\phi_1$ & 18.504 & 1.000 & 1.381 & 0.228 & 0.202 &  & 0.170 & 0.131 \\
$\sigma_1^2$ & 16.773 & 1.000 & 1.528 & 0.131 & 0.098 &  & 0.051 & 0.039 \\
$c_2$ & 17.712 & 1.000 & 1.686 & 0.145 & 0.083 &  & 0.082 & 0.062 \\
$\phi_2$ & 14.109 & 1.000 & 1.442 & 0.088 & 0.076 &  & 0.068 & 0.051 \\
$\sigma_2^2$ & 12.829 & 1.000 & 1.616 & 0.082 & 0.056 &  & 0.051 & 0.046 \\
$c_3$ & 15.212 & 1.000 & 1.517 & 0.182 & 0.087 &  & 0.091 & 0.087 \\
$\phi_3$ & 25.558 & 1.000 & 1.111 & 0.125 & 0.106 &  & 0.077 & 0.062 \\
$\sigma_3^2$ & 26.736 & 1.000 & 0.875 & 0.116 & 0.107 &  & 0.066 & 0.050 \\
 &  &  &  &  &  &  &  &  \\
Marg. Lik. & 6.720 & 1.000 & 1.075 & 0.042 & 0.032 &  & 0.033 & 0.018 \\
 &  &  &  &  &  &  &  &  \\
Time (s) & 5603 & 285 & 430 & 154 & 194 &  & 167 & 204 \\
\hline\hline
\end{tabular}
\end{threeparttable}}
\end{center}
\end{table}

\subsection{Particle Marginal Metropolis-Hastings}

We now consider the use of the bootstrap filter, EIS and particle EIS algorithms for the PMMH estimation of the posterior distribution of the bivariate stochastic volatility model. We implement two Metropolis-Hastings (M-H) algorithms: the adaptive random walk method of \citeN{rr2009} and the independent M-H method using the MitISEM proposal.  We run 50,000 iterations of the two algorithms and discard a burn-in sample of 5,000 iterations.

Table \ref{tab:example6} reports the acceptance rates, the inefficiency factors (calculated using the overlapping batch means method), and the total computing times in hours. The results show that only the independent Metropolis-Hastings using the MitISEM proposal in combination with the particle EIS method for estimating the likelihood performs satisfactorily. The particle EIS method achieves an acceptance rate of 0.423 and inefficiency factors between 7.3 and 16.4 when using the MitISEM (EIS) proposal, in a total computing time of 4.8 hours.  That compares to acceptance rates of 0.119 and 0.199 and computing times of 79.5 and 8.2 hours for the BF and EIS methods respectively, with inefficiency factors higher than 40 for all the parameters.  We also find that the MitISEM (P-EIS) proposal leads to an increase in the acceptance rate to 0.588 and substantial improvements in the inefficiency factors.

\begin{table}[!tbp]
\begin{center}
{\small \caption{{\sc Bivariate SV - PMMH inefficiencies under different likelihood estimation methods}.}\label{tab:example6}
\vspace{-12pt}
\begin{threeparttable}
{\footnotesize The table examines the performance of different particle marginal Metropolis-Hastings (PMMH) methods for estimating the posterior distribution of the bivariate stochastic volatility model. The table reports the acceptance rates, the inefficiency factors for each parameter, and the total computing time in hours. }
\vspace{7pt}
\begin{tabular}{lccccccccc}
\hline\hline
\multicolumn{1}{c}{} & \multicolumn{ 3}{c}{Adaptive Random Walk} & \multicolumn{1}{l}{} & \multicolumn{ 3}{c}{MitISEM (EIS)} & \multicolumn{1}{l}{} & \multicolumn{1}{l}{MitISEM (P-EIS)} \\
\multicolumn{1}{c}{} & BF & EIS & P-EIS &  & BF & EIS & P-EIS &  & P-EIS \\
\multicolumn{1}{c}{} & N=5,000 & N=300 & N=50 &  & N=5,000 & N=300 & N=50 &  & N=50 \\
\cline{2-4}\cline{6-8}\cline{10-10}
Acc. rate & 0.082 & 0.116 & 0.241 &  & 0.119 & 0.199 & 0.423 &  & 0.588 \\
\\
$c_1$ & 107.4 & 94.3 & 37.1 &  & 44.2 & 60.4 & 16.4 &  & 11.1 \\
$\phi_1$ & 96.4 & 89.4 & 32.9 &  & 62.8 & 62.6 & 9.5 &  & 6.0 \\
$\sigma_1^2$ & 82.6 & 83.8 & 31.4 &  & 54.5 & 68.8 & 7.0 &  & 4.7 \\
$c_2$ & 91.9 & 97.3 & 31.9 &  & 43.1 & 47.7 & 13.3 &  & 7.1 \\
$\phi_2$ & 85.9 & 78.8 & 32.3 &  & 53.0 & 87.3 & 7.5 &  & 4.2 \\
$\sigma_2^2$ & 85.6 & 85.4 & 32.7 &  & 56.7 & 69.1 & 8.4 &  & 5.7 \\
$c_3$ & 99.0 & 89.3 & 32.9 &  & 52.5 & 63.7 & 7.3 &  & 4.7 \\
$\phi_3$ & 85.8 & 99.7 & 33.1 &  & 43.4 & 75.6 & 10.0 &  & 5.4 \\
$\sigma_3^2$ & 77.6 & 100.2 & 32.9 &  & 48.5 & 74.3 & 9.8 &  & 5.0 \\
\\
Time (h) & 79.6 & 8.5 & 5.1 &  & 79.5 & 8.2 & 4.8 &  & 4.4 \\
\hline\hline
\end{tabular}
\end{threeparttable}}
\end{center}
\end{table}


\newpage

\bibliographystyle{ecta}
\bibliography{references}

\begin{thebibliography}{34}
\newcommand{\enquote}[1]{``#1''}
\expandafter\ifx\csname natexlab\endcsname\relax\def\natexlab#1{#1}\fi

\bibitem[\protect\citeauthoryear{Andrieu, Doucet, and Holenstein}{Andrieu
  et~al.}{2010}]{pmcmc}
\textsc{Andrieu, C., A.~Doucet, and R.~Holenstein} (2010): \enquote{{Particle
  Markov chain Monte Carlo methods},} \emph{Journal of the Royal Statistical
  Society - Series B: Statistical Methodology}, 72, 269--342.

\bibitem[\protect\citeauthoryear{Barra, Hoogerheide, Koopman, and Lucas}{Barra
  et~al.}{2013}]{bhkl2013}
\textsc{Barra, I., L.~Hoogerheide, S.~J. Koopman, and A.~Lucas} (2013):
  \enquote{{Joint independent Metropolis-Hastings methods for nonlinear
  non-Gaussian state space models},} Working paper, Tinbegen Institute.

\bibitem[\protect\citeauthoryear{Bauwens and Galli}{Bauwens and
  Galli}{2009}]{BG09}
\textsc{Bauwens, L. and F.~Galli} (2009): \enquote{Efficient Importance
  Sampling for ML Estimation of SCD Models,} \emph{Computational Statistics and
  Data Analysis}, 53, 1974--1992.

\bibitem[\protect\citeauthoryear{Chopin}{Chopin}{2004}]{chopin2004}
\textsc{Chopin, N.} (2004): \enquote{{Central limit theorem for sequential
  Monte Carlo methods and its application to Bayesian inference},} \emph{The
  Annals of Statistics}, 32, 2385--2411.

\bibitem[\protect\citeauthoryear{DeJong, Liesenfeld, Moura, Richard, and
  Dharmarajan}{DeJong et~al.}{2012}]{dlmrd2012}
\textsc{DeJong, D.~N., R.~Liesenfeld, G.~V. Moura, J.-F. Richard, and
  H.~Dharmarajan} (2012): \enquote{{Efficient likelihood evaluation of
  state-space representations},} \emph{The Review of Economic Studies}, 80,
  538--567.

\bibitem[\protect\citeauthoryear{{Del Moral}}{{Del Moral}}{2004}]{delmoral2004}
\textsc{{Del Moral}, P.} (2004): \emph{{Feynman-Kac formulae: genealogical and
  interacting particle systems with applications}}, New York: Springer.

\bibitem[\protect\citeauthoryear{Doucet, Briers, and S\'{e}n\'{e}cal}{Doucet
  et~al.}{2006}]{dbs2006}
\textsc{Doucet, A., M.~Briers, and S.~S\'{e}n\'{e}cal} (2006):
  \enquote{{Efficient Block Sampling Strategies for Sequential Monte Carlo
  Methods},} \emph{Journal of Computational and Graphical Statistics}, 15,
  693--711.

\bibitem[\protect\citeauthoryear{Durbin and Koopman}{Durbin and
  Koopman}{1997}]{DK97}
\textsc{Durbin, J. and S.~J. Koopman} (1997): \enquote{Monte Carlo Maximum
  Likelihood Estimation for non-Gaussian State Space Models,}
  \emph{Biometrika}, 669--684.

\bibitem[\protect\citeauthoryear{Durbin and Koopman}{Durbin and
  Koopman}{2000}]{DK00}
---\hspace{-.1pt}---\hspace{-.1pt}--- (2000): \enquote{Time series analysis of
  non-Gaussian observations based on state space models from both classical and
  Bayesian perspectives,} \emph{Journal of the Royal Statistical Society,
  Series B}, 3--56.

\bibitem[\protect\citeauthoryear{Durbin and Koopman}{Durbin and
  Koopman}{2001}]{DK1}
---\hspace{-.1pt}---\hspace{-.1pt}--- (2001): \emph{Time Series Analysis by
  State Space Methods}, Oxford University Press.

\bibitem[\protect\citeauthoryear{Fearnhead, Wyncoll, and Tawn}{Fearnhead
  et~al.}{2010}]{fwt2010}
\textsc{Fearnhead, P., D.~Wyncoll, and J.~Tawn} (2010): \enquote{{A sequential
  smoothing algorithm with linear computational cost},} \emph{Biometrika}, 97,
  447--464.

\bibitem[\protect\citeauthoryear{Flury and Shephard}{Flury and
  Shephard}{2011}]{fs2011}
\textsc{Flury, T. and N.~Shephard} (2011): \enquote{{Bayesian inference based
  only on simulated likelihood: particle filter analysis of dynamic economic
  models},} \emph{Econometric Theory}, 27, 933--956.

\bibitem[\protect\citeauthoryear{Geweke}{Geweke}{1989}]{Geweke89}
\textsc{Geweke, J.} (1989): \enquote{Bayesian Inference in Econometric Models
  Using Monte Carlo Integration,} \emph{Econometrica}, 57, 1317--1739.

\bibitem[\protect\citeauthoryear{Godsill, Doucet, and West}{Godsill
  et~al.}{2004}]{gdw2004}
\textsc{Godsill, S.~J., A.~Doucet, and M.~West} (2004): \enquote{{Monte Carlo
  smoothing for nonlinear time series},} \emph{Journal of the American
  Statistical Association}, 99, 156--168.

\bibitem[\protect\citeauthoryear{Hafner and Manner}{Hafner and
  Manner}{2012}]{hm2012}
\textsc{Hafner, C.~M. and H.~Manner} (2012): \enquote{{Dynamic stochastic
  copula models: estimation, inference and applications},} \emph{Journal of
  Applied Econometrics}, 27, 269--295.

\bibitem[\protect\citeauthoryear{Harvey, Ruiz, and Shephard}{Harvey
  et~al.}{1994}]{Harvey94}
\textsc{Harvey, A., E.~Ruiz, and N.~Shephard} (1994): \enquote{Multivariate
  Stochastic Variance Models,} \emph{Review of Economic Studies}, 61, 247--264.

\bibitem[\protect\citeauthoryear{Hoogerheide, Opschoor, and van
  Dijk}{Hoogerheide et~al.}{2012}]{hod2012}
\textsc{Hoogerheide, L., A.~Opschoor, and H.~K. van Dijk} (2012): \enquote{{A
  class of adaptive importance sampling weighted EM algorithms for efficient
  and robust posterior and predictive simulation},} \emph{Journal of
  Econometrics}, 171, 101--–120.

\bibitem[\protect\citeauthoryear{Kitagawa}{Kitagawa}{1996}]{kitagawa1996}
\textsc{Kitagawa, G.} (1996): \enquote{{Monte Carlo filter and smoother for
  non-Gaussian nonlinear state space models},} \emph{Journal of Computational
  and Graphical Statistics}, 5, 1--25.

\bibitem[\protect\citeauthoryear{Kleppe and Liesenfeld}{Kleppe and
  Liesenfeld}{2013}]{kl2013}
\textsc{Kleppe, T.~S. and R.~Liesenfeld} (2013): \enquote{{Efficient importance
  sampling in mixture frameworks},} \emph{Computational Statistics \& Data
  Analysis}, forthcoming.

\bibitem[\protect\citeauthoryear{Koopman, Lucas, and Scharth}{Koopman
  et~al.}{2012}]{NAIS}
\textsc{Koopman, S.~J., A.~Lucas, and M.~Scharth} (2012): \enquote{{Numerically
  accelerated importance sampling for nonlinear non-Gaussian state space
  models},} Working paper, Tinbergen Institute.

\bibitem[\protect\citeauthoryear{Koopman, Lucas, and Scharth}{Koopman
  et~al.}{2013}]{KLS3}
---\hspace{-.1pt}---\hspace{-.1pt}--- (2013): \enquote{Static and dynamic
  multivariate Gaussian efficient importance sampling,} Working paper.

\bibitem[\protect\citeauthoryear{Koopman, Shephard, and Creal}{Koopman
  et~al.}{2009}]{KSC09}
\textsc{Koopman, S.~J., N.~Shephard, and D.~Creal} (2009): \enquote{Testing the
  assumptions behind importance sampling,} \emph{Journal of Econometrics}, 149,
  2--11.

\bibitem[\protect\citeauthoryear{Liesenfeld and Richard}{Liesenfeld and
  Richard}{2003}]{LR2003}
\textsc{Liesenfeld, R. and J.-F. Richard} (2003): \enquote{Univariate and
  Multivariate Stochastic Volatility Models: Estimation and Diagnostics,}
  \emph{Journal of Empirical Finance}, 10, 505--531.

\bibitem[\protect\citeauthoryear{Liesenfeld and Richard}{Liesenfeld and
  Richard}{2010}]{lr2010}
---\hspace{-.1pt}---\hspace{-.1pt}--- (2010): \enquote{{Efficient estimation of
  probit models with correlated errors},} \emph{Journal of Econometrics}, 156,
  367--376.

\bibitem[\protect\citeauthoryear{Liesenfeld, Richard, and Vogler}{Liesenfeld
  et~al.}{2013}]{lrv2013}
\textsc{Liesenfeld, R., J.-F. Richard, and J.~Vogler} (2013):
  \enquote{{Analysis of discrete dependent variable models with spatial
  correlation},} Working paper.

\bibitem[\protect\citeauthoryear{Lin, Chen, and Liu}{Lin
  et~al.}{2013}]{lcl2013}
\textsc{Lin, M., R.~Chen, and J.~S. Liu} (2013): \enquote{{Lookahead Strategies
  for Sequential Monte Carlo},} \emph{Statistical Science}, 28, 69--94.

\bibitem[\protect\citeauthoryear{Liu and Chen}{Liu and Chen}{1998}]{lc1998}
\textsc{Liu, J.~S. and R.~Chen} (1998): \enquote{{Sequential Monte Carlo
  methods for dynamic systems},} \emph{Journal of the American Statistical
  Association}, 93, 1032--1044.

\bibitem[\protect\citeauthoryear{Pitt}{Pitt}{2002}]{pitt2002}
\textsc{Pitt, M.~K.} (2002): \enquote{{Smooth particle filters for likelihood
  evaluation and maximisation},} Working paper.

\bibitem[\protect\citeauthoryear{Pitt and Shephard}{Pitt and
  Shephard}{1999}]{ps1999}
\textsc{Pitt, M.~K. and N.~Shephard} (1999): \enquote{{Filtering via
  simulation: auxiliary particle filters},} \emph{Journal of the American
  Statistical Association}, 94, 590--599.

\bibitem[\protect\citeauthoryear{Pitt, Silva, Giordani, and Kohn}{Pitt
  et~al.}{2012}]{psgk2012}
\textsc{Pitt, M.~K., R.~d.~S. Silva, P.~Giordani, and R.~Kohn} (2012):
  \enquote{{On some properties of Markov chain Monte Carlo simulation methods
  based on the particle filter},} \emph{Journal of Econometrics}, 171,
  134--151.

\bibitem[\protect\citeauthoryear{Richard and Zhang}{Richard and
  Zhang}{2007}]{ZR2007}
\textsc{Richard, J.-F. and W.~Zhang} (2007): \enquote{Efficient
  High-Dimensional Importance Sampling,} \emph{Journal of Econometrics}, 141,
  1385--1411.

\bibitem[\protect\citeauthoryear{Roberts and Rosenthal}{Roberts and
  Rosenthal}{2009}]{rr2009}
\textsc{Roberts, G.~O. and J.~S. Rosenthal} (2009): \enquote{{Examples of
  adaptive MCMC},} \emph{Journal of Computational and Graphical Statistics},
  18, 349--367.

\bibitem[\protect\citeauthoryear{Shephard and Pitt}{Shephard and
  Pitt}{1997}]{SP97}
\textsc{Shephard, N. and M.~Pitt} (1997): \enquote{Likelihood analysis of
  non-Gaussian measurement time series,} \emph{Biometrika}, 84, 653--667.

\bibitem[\protect\citeauthoryear{Tran, Scharth, Pitt, and Kohn}{Tran
  et~al.}{2013}]{tspk2013}
\textsc{Tran, M.-N., M.~Scharth, M.~K. Pitt, and R.~Kohn} (2013):
  \enquote{Importance Sampling Squared for Bayesian Inference in Latent
  Variable Models,} Mimeo.

\end{thebibliography}

\newpage

\appendix
\section*{Appendix}
\section{EIS for state space models with nonlinear transition and additive Student's $t$ state disturbances}\label{sec:EISsampler}

This appendix develops an efficient importance sampling method for a state space model with nonlinear transition and additive Student's $t$ state disturbances which includes the univariate SV models of Section \ref{sec:models} as special cases. The method follows from \citeN{kl2013}, which consider the case in which the measurement density $p(y_t|x_t)$ is a continuous or discrete mixture. They propose a data augmentation scheme in which they explicitly include the mixture components in the integrand of \eqref{lik}. This allows them to approximate the different components of the measurement density separately using the EIS method.

Applying this principle to our current setting, we consider the modified transition density $p(x_t|x_{t-1},\lambda_t)p(\lambda_t)$, where $p(x_t|x_{t-1},\lambda_t)$ is a Gaussian density and $\lambda_t$ is a vector of inverse gamma random variables. The state space model is
\begin{align*}
&y_t|x_t\sim p(y_{t}|Zx_t),\qquad x_{t}=F(x_{t-1})+\Lambda_t \eta_{t},\qquad x_1\sim \Nn(a_1,P_1), \qquad \eta_{t}\sim \Nn(0,Q),
\end{align*}
where $y_t$ is the observation vector, $x_t$ is the $m\times1$ state vector, $Z$ is a $p\times m$ (with $p\leq m$) and $F(.)$ is a $\mathbb{R}^m\rightarrow\mathbb{R}^m$ nonlinear function.  The scaling matrix $\Lambda_t$ is diagonal with entries $\sqrt{\lambda_{1,t}},\ldots,\sqrt{\lambda_{m,t}}$, where $\lambda_{j,t} \sim \textrm{IG}(\nu_{\eta,j}/2,\nu_{\eta,j}/2)$. All the random variables $\lambda_{j,t}$ are mutually independent. We have that $\lambda_{t}=(\lambda_{1,t} \, , \, \ldots \, , \, \lambda_{m,t})'$. We write the measurement density in terms of the signal vector $Zx_t$ instead of the state vector $x_t$ in order to reduce the computational cost of running the EIS algorithm when $p<m$, see for example \citeN{NAIS}.

After data augmentation, the likelihood function \eqref{lik} becomes
\begin{align}
    &=\int p(y_t|x_1)p(x_1)\prod_{t=2}^{n} p(y_t|x_t)p(x_t|x_{t-1},\lambda_t)p(\lambda_t)\dd x_1\ldots \dd x_n\,\dd \lambda_2\ldots \dd \lambda_n.\nonumber
\end{align}

We consider the sequential importance densities
\begin{equation*}
q(x_t,\lambda_t|x_{t-1},y_{1:n})=q(x_t|x_{t-1},\lambda_t,y_{1:n})q(\lambda_t|y_{1:n}),
\end{equation*}
where
\begin{equation}\label{eq:idgaussian}
q(x_t|x_{t-1},\lambda_t,y_{1:n})=\delta_{t}(x_{t-1},\lambda_t)\exp \left(b_t^{\prime} \, Zx _t-\frac 12 x_t^{\prime}Z^{\prime}\, C_t \, Zx_t \right)p(x_t|x_{t-1},\lambda_t)
\end{equation}
and
\begin{equation}\label{eq:idig}
q(\lambda_t|y_{1:n})=\(\prod_{j=1}^{m}\varphi_{j,t}\lambda_{j,t}^{\alpha_{j,t}}\exp(\beta_{j,t}/\lambda_{j,t})\)p(\lambda_t).
\end{equation}
The importance parameters are $b_t$, $C_t$, $\alpha_{1,t},\ldots,\alpha_{m,t}$ and $\beta_{1,t},\ldots,\beta_{m,t}$ . The terms $\delta_{t}(x_{t-1},\lambda_t)$ and $\varphi_{j,t}$ are constants that ensure that $q(x_t,\lambda_t|x_{t-1},y_{1:n})$ integrates to one.

The importance densities in \eqref{eq:idgaussian} and \eqref{eq:idig} offset the model transition densities and use conjugate terms to approximate the measurement densities and integration constants. With some algebra, we can show that $q(x_t|x_{t-1},\lambda_t,y_{1:n})$ is a Gaussian density with covariance matrix
\begin{equation*}\label{eq:EISmean}
V_t=[(\Lambda_tQ\Lambda_t)^{-1}+C_t]^{-1}
\end{equation*}
and mean vector
\begin{equation*}\label{eq:EISvar}
\mu_t=V_t[Z'b_t+(\Lambda_tQ\Lambda_t)^{-1}F(x_{t-1})],
\end{equation*}
while the importance density $q(\lambda_t|y_{1:n})$ is such that
\begin{equation*}
\lambda_{j,t}\sim \textrm{IG}(\nu_{\eta,j}/2-\alpha_{j,t},\nu_{\eta,j}/2-\beta_{j,t})
\end{equation*}
for $j=1,\ldots,m$. The constants are
\begin{equation*}\label{eq:intconst1}
\log \delta_{t}(x_{t-1},\lambda_t)=\frac{1}{2}\log(|\Lambda_tQ\Lambda_t|/|V_t|)+\frac{1}{2}F(x_{t-1})'(\Lambda_tQ\Lambda_t)^{-1}F(x_{t-1})-\frac{1}{2}\mu_t'V_t^{-1}\mu_t
\end{equation*}
and
\begin{equation*}
\varphi_{j,t}=\frac{\Gamma(\nu_{\eta,j}/2)}{(\nu_{\eta,j}/2)^{\nu_{\eta,j}/2}}\frac{(\nu_{\eta,j}/2-\beta_{j,t})^{\nu_{\eta,j}/2-\alpha_{j,t}}}{\Gamma(\nu_{\eta,j}/2-\beta_{j,t})}.
\end{equation*}

To implement Algorithm 1, suppose we generate draws $x^{(1)},\ldots,x^{(S)},\lambda^{(1)},\ldots,\lambda^{(S)}$ from the current candidate density $q^{[k]}(x_{1:n},\lambda_{2:n}|y_{1:n})$. Following an appropriate modification of \eqref{eq:eisreg} and \eqref{eq:mincrit} for the data augmentation setting, we update the importance parameters by running backwards recursively for every period $t$ ordinary least squared regressions with dependent variable
\begin{equation*}
\log p(y_t|x_t^{(s)})-\log\delta_{t+1}(x_{t}^{(s)},\lambda_{t+1}^{(s)})
\end{equation*}
and regressors
\begin{align*}
&Zx_t^{(s)},\,\,-(1/2)\textrm{vech}[(Zx_t^{(s)})(Zx_t^{(s)})^\prime],\nonumber\\
&\log(\lambda_{1,t+1}^{(s)}),\ldots \log(\lambda_{m,t+1}^{(s)}),\,1/\lambda_{1,t+1}^{(s)},\ldots, 1/\lambda_{m,t+1}^{(s)}
\end{align*}
plus a constant. We need to multiply the coefficients associated with the off-diagonal elements of $-(1/2)\textrm{vech}[(Zx_t^{(s)})(Zx_t^{(s)})^\prime]$ by two because these terms appear twice in the quadratic form in \eqref{eq:idgaussian}. The resulting coefficients after these steps give us $b_t$, $C_t$, $\alpha_{1,t+1},\ldots,\alpha_{m,t+1}$ and $\beta_{1,t+1},\ldots,\beta_{m,t+1}$ respectively. Because the inverse gamma variables appear directly in $\log\delta_{t}(x_{t-1}^{(s)},\lambda_{t}^{(s)})$ and not $p(y_t|x_t^{(s)})$, it is necessary to estimate the coefficients of $q(\lambda_t|y_{1:n})$ jointly with
$q(x_{t-1}|x_{t-2},\lambda_{t-1},y_{1:n})$. Though this may initially seem counterintuitive, the need for this design highlights the importance of the integration constants in the EIS method.

We emphasised that Algorithm 1 is based on common random numbers (CRNs). The use of CRNs for the current problem requires computationally expensive inversions of gamma cumulative density functions. We circumvent this issue by first fixing $q(\lambda_t|y_{1:n})=p(\lambda_t)$ and letting the importance parameters $b_t$ and $C_t$ converge for $t=1,\ldots,n$. That provides the partial EIS density which we use in Section \ref{sec:simresults}. We then run only one iteration of the full EIS regressions described above using the partial EIS parameters as starting values. We have found that additional iterations generate modest gains in efficiency that do not compensate for the added computational cost when using CRNs.

\end{document}